\documentclass[aps,twocolumn,prx,preprintnumbers,showpacs,amsmath,amssymb,superscriptaddress]{revtex4-1}
\usepackage[colorlinks, citecolor=blue]{hyperref}
\usepackage{graphicx}
\usepackage{dcolumn}
\usepackage{threeparttable}
\usepackage{multirow}
\usepackage{booktabs}
\usepackage{txfonts}
\usepackage{xcolor}
\usepackage{bm}
\usepackage{amssymb}
\usepackage{amsmath}
\usepackage{latexsym}
\usepackage{epsfig}
\usepackage{amsbsy}
\usepackage{array}
\usepackage{tabularx}

\newcommand{\ii}{\mathrm{i}}
\newcommand{\ee}{\mathrm{e}}

\newcommand{\qq}{\mathbf{q}}
\newcommand{\kk}{\mathbf{k}}
\newcommand{\nk}{{n\mathbf{\kk}}}
\newcommand{\mk}{{m\mathbf{\kk}}}

\newcommand{\sg}{\mathbf{g}}

\newcommand{\lG}{\mathbf{G}}

\newcommand{\UG}{\overline{G}}
\newcommand{\OG}{\underline{G}}

\begin{document}

\title{Beyond the quasiparticle approximation: Fully self-consistent $GW$ calculations}

\author{Manuel Grumet}
\affiliation{University of Vienna, Faculty of Physics and Center for
Computational Materials Science, Sensengasse 8/12, A-1090 Vienna,
Austria}

\author{Peitao Liu}
\email{peitao.liu@univie.ac.at}
\affiliation{University of Vienna, Faculty of Physics and Center for
Computational Materials Science, Sensengasse 8/12, A-1090 Vienna,
Austria}

\author{Merzuk Kaltak}
\affiliation{University of Vienna, Faculty of Physics and Center for
Computational Materials Science, Sensengasse 8/12, A-1090 Vienna,
Austria}

\author{Ji\v{r}\'{\i} Klime\v{s}}
\affiliation{J. Heyrovsk\'{y} Institute of Physical Chemistry, Academy
of Sciences of the Czech Republic, Dolej\v{s}kova 3, CZ-18223 Prague 8,
Czech Republic}
\affiliation{Department of Chemical Physics and Optics, Faculty of
Mathematics and Physics, Charles University, Ke Karlovu 3, CZ-12116
Prague 2, Czech Republic}

\author{Georg Kresse}
\email{georg.kresse@univie.ac.at}
\affiliation{University of Vienna, Faculty of Physics and Center for
Computational Materials Science, Sensengasse 8/12, A-1090 Vienna,
Austria}

\begin{abstract}
We present quasiparticle (QP) energies from fully
self-consistent $GW$ (sc$GW$) calculations for a set of prototypical semiconductors
and insulators within the framework of the projector-augmented wave
methodology.  To obtain converged results, both finite
basis-set corrections and $k$-point corrections are included, and a simple procedure
is suggested to deal with the singularity of the Coulomb kernel in the long-wavelength limit, the so called
head correction. It
is shown that the inclusion of the head corrections in the sc$GW$
calculations is critical to obtain accurate QP energies with a reasonable $k$-point set.
We first validate our implementation by presenting detailed results for the selected case of
diamond, and then we  discuss the converged QP energies, in particular the
band gaps, for a set of gapped compounds and compare them to
single-shot $G_0W_0$, QP self-consistent $GW$, and previously available
sc$GW$ results as well as experimental results.
\end{abstract}

\maketitle

%--------------------------------------------------------------------------------
\section{Introduction}\label{sec:intro}
%--------------------------------------------------------------------------------

Hedin's equations~\cite{Hedin1965,Hedin1969} are in principle a rigorous
and exact way to calculate quasiparticle (QP) energies (electron
addition or removal energies).
Nevertheless, in practice the equations cannot be
solved exactly due to the need to perform the
calculations self-consistently and
difficulties in including the vertex correction, which is defined as
the functional derivative of the self-energy with respect to the external potential.
Therefore, approximations are strictly required. One
of the most widely used approximations is the $GW$
approximation~\cite{Hedin1965}, which neglects the vertex completely. Related to
this, it is furthermore common to start from orbitals determined using
density functional theory (DFT) and to perform so-called single-shot
$G_0W_0$ calculations~\cite{Strinati1980, Strinati1982, Louie1985,
Louie1986}. This generally gives good agreement with experiments for
extended, moderately correlated materials because of a cancellation of errors originating from
the lack of self-consistency and the absence of vertex corrections.
In order to go beyond $G_0W_0$, several strategies such as the cumulant
expansion~\cite{aryasetiawan:1996, Guzzo_CE2011, kas:2014, Caruso2015,
gumhalter:2016}, inclusion of some approximate vertex~\cite{kutepov:2016,
kutepov:2017, grueneis:2014, maggio:2017a}, and quantum chemistry methods like
algebraic-diagrammatic construction (ADC) and equation-of-motion
coupled-cluster~\cite{mcclain:2017} have been proposed.

A problem that most single-shot Green's function based methods have in common is that
some conservation laws such as energy and particle number conservation~\cite{Baym1961,Baym1962} are violated.
Due to their perturbative nature, the results also depend on the starting one-electron energies and orbitals,
which are usually obtained from the solution of the Kohn-Sham (KS)
equations or generalized KS schemes~\cite{fuchs:2007, bruneval:2013}.
This issue can be avoided by performing the calculations self-consistently.
Eigenvalue self-consistent $GW$~\cite{Zhu1991,
Zakharov1994, Shishkin_PRB2007}, which updates the eigenvalues either
only in the Green's functions $G$ (ev-$GW_0$), or both in $G$ and the
screened interactions $W$ (ev-$GW$), while the orbitals remain fixed,
generally improves the description of band gaps towards the experimental
values as compared to $G_0W_0$. Quasiparticle self-consistent $GW$
(QP$GW$) removes the starting-point dependence entirely by determining an
optimum effective non-local static exchange-correlation potential~\cite{Faleev2004,schilfgaarde:2006,
bruneval:2006,shishkin:2007a}. However, it overestimates band gaps in
solids due to the underestimation of the dielectric screening in the
random phase approximation (RPA)~\cite{shishkin:2007a,Bhandari2018}.
Self-consistent $GW$ (sc$GW$) avoids the quasiparticle approximation, and
the Dyson equation for the Green's function and $W$ are solved fully
self-consistently~\cite{Holm-scGW1998,schoene:1998, Adrian2009,
caruso:2012,Kutepov2012,Koval2014, kutepov:2016, kutepov:2017}.
In addition, the self-consistent $W$ is invariant under spatial and time translations,
so conservation laws (momentum, energy and particle number conservation)~\cite{Baym1961,Baym1962}
are satisfied in sc$GW$.
Nevertheless, without vertex corrections, sc$GW$ shows a significant
overestimation of the bandwidth for metals and band gaps for gapped
systems. Recently, there have been attempts to include the vertex in $W$
and the self-energy for crystalline materials~\cite{kutepov:2016,
kutepov:2017, grueneis:2014, maggio:2017a}, showing a substantial
improvement on the bandwidths, ionization potentials and band gaps
compared to sc$GW$. These approaches are computationally exceedingly
demanding and will not be considered in the present work.

Although there are already some studies that are dedicated to full
sc$GW$ calculations~\cite{Holm-scGW1998,schoene:1998, Adrian2009,
caruso:2012,Kutepov2012,Koval2014, kutepov:2016, kutepov:2017}, they are
restricted to few systems and reference results for a more extensive set
of materials are still missing. The main problems in obtaining reference
values for solids are threefold. First, full sc$GW$ calculations are
technically demanding. Second, the basis set convergence for
the QP energies is very slow~\cite{Gulans2014, klimes:2014, ergoenenc:2018}. Third,
there is a singularity problem associated with the long-wavelength limit
of the product of the Coulomb kernel and the dielectric function. Within $G_0W_0$, this issue can
be solved straightforwardly using $\kk\cdot\mathbf{p}$ perturbation
theory~\cite{Baroni1986,gajdos:2006, harl:2010}, but this is intractable
for sc$GW$.

The goal of this paper is to obtain converged sc$GW$ QP energies for a
set of semiconductors and insulators within the framework of
the projector-augmented wave (PAW) methodology. To establish reference values, we include
finite basis set corrections, as well as $k$-point corrections. The
singularity problem in the product of the Coulomb kernel and the dielectric matrix
is overcome by an extrapolation from the available
results at finite $\mathbf{q}$. In addition to sc$GW$ results, we also report results for $G_0W_0$ and
QP$GW$ calculations. It should be noted
that in the present work, vertex corrections are not considered and
therefore it is expected that our sc$GW$ results will overestimate the band gaps
as compared to the experimental values.

The paper is organized as follows. In Sec.~\ref{sec:method} we will
detail the methodology of our sc$GW$ implementation. Particular emphasis
is devoted to the extrapolation scheme that is used to solve the
singularity problem of the Coulomb operator. Technical
details and computational setups will be provided in
Sec.~\ref{sec:details}. The results will be presented and discussed in
Sec.~\ref{sec:results} and summarized in Sec.~\ref{sec:conlcusions}.

%--------------------------------------------------------------------------------
\section{Method}\label{sec:method}
%--------------------------------------------------------------------------------

%--------------------------------------------------------------------------------
\subsection{Self-consistent $GW$}\label{sec:scgw}
%--------------------------------------------------------------------------------

Our sc$GW$ scheme is based on our recent cubic-scaling $GW$
implementation~\cite{liu:2016}, where the polarizability and self-energy
are calculated in the real-space and imaginary-time
domains~\cite{Rojas1995,rieger:1999}. Efficient temporal discrete Fourier
transformations with only a few nonuniform optimized imaginary grid
points~\cite{KaltakJCTC2014} and spatial fast Fourier transformations
(FFT)~\cite{Kaltak2014} allow for fast QP calculations with a scaling
that is cubic in the system size and linear in the number of $k$-points that are
used to sample the Brillouin zone. The implementation has been validated
by successfully predicting QP energies of typical semiconductors,
insulators and metals as well as
molecules~\cite{liu:2016, MaggioGW1002017,Tomczak2017}. Here, we go one step further and
introduce self-consistency both in $G$ and $W$. It needs to be mentioned
that, for consistency, in this paper we follow almost the same notations
and definitions that were used in our previous
publication~\cite{liu:2016}. In the following, we present our sc$GW$
implementation in detail.

Starting from the correlated self-energy $\Sigma^\mathrm{c}(\ii\omega)$
obtained from $G_0W_0$~\cite{liu:2016}, the new interacting Green's
function $G(\ii\omega)$ for the next iteration is calculated in the
Hartree-Fock (HF) canonical-orbital basis by the  Dyson's equation
\begin{equation}
G(\ii\omega) =
\left[ \ii\omega+\mu - H^\mathrm{HF} - \Sigma^\mathrm{c}(\ii\omega) \right]^{-1},
\end{equation}
where $\mu$ is the Fermi energy and
$H^\mathrm{HF} = T + V_{n-e} + V_H + \Sigma^\mathrm{x}$ is the
HF Hamiltonian, with $T$, $ V_{n-e}$, $V_H$ and
$\Sigma^\mathrm{x}$ being the kinetic energy, the potential from the
nuclei, the Hartree potential and the exact exchange, respectively.
Note that in the present work, $\mu$ is always set to the HF Fermi energy,
which is located at mid gap between the HF valence band maximum  (VBM) and conduction band minimum (CBM).
This does not introduce any approximation, since in sc$GW$ one can chose the Fermi-level anywhere between the
sc$GW$ VBM and CBM, which are not broadened by lifetime effects.
For the materials considered in this work, this was always the case.

The interacting density matrix is then calculated in the canonical HF basis by
\begin{eqnarray}
\label{eq:definition_density_matrix}
\Gamma_{ij} = \frac{1}{2\pi}
  \int_{-\infty}^{+\infty} d\omega \, G_{ij}(\ii\omega).
\end{eqnarray}
However, this integral usually diverges. To address this issue, $G(\ii\omega)$
is split into two parts
\begin{eqnarray}
\label{eq:split_G_into_two}
G(\ii\omega)=G^{\rm HF}(\ii\omega)+G^c(\ii\omega).
\end{eqnarray}
Here, $G^{\rm HF}(\ii\omega)$ is the HF Green's function
\begin{eqnarray}
\label{eq:HF_G0_1}
G^{\rm HF}(\ii\omega)=[\ii\omega+\mu- H^{\rm HF}]^{-1},
\end{eqnarray}
and $G^c(\ii\omega)$ is the correlated part of the Green's function. Due
to the splitting in Eq. (\ref{eq:split_G_into_two}), the density matrix
includes two contributions
\begin{eqnarray}\label{eq:density_split_into_two}
\Gamma_{ij}=\Gamma_{ij}^{\rm HF}+\Gamma_{ij}^c,
\end{eqnarray}
where the calculation of the HF density matrix $\Gamma_{ij}^{\rm HF}$ is
straightforward,
\begin{eqnarray}\label{eq:density_HF}
\Gamma_{ij}^{\rm HF} = \theta\left(\mu-\epsilon^{\rm HF}_{i}\right) \, \delta_{ij}.
\end{eqnarray}
Here, $\theta$ is the Heaviside step function and $\epsilon^{\rm HF}_{i}$ are the eigenvalues of the HF Hamiltonian
(we note again that the matrices are presented in the canonical HF basis, making
$H^{\rm HF}(\ii\omega)$ and $G^{\rm HF}(\ii\omega)$ diagonal).
Since $\Sigma^\mathrm{c}(\ii\omega)$ decays as
$1/(\ii\omega)$~\cite{wang:2011}, $G^c(\ii\omega)$ decays as
$1/(\ii\omega)^3$. The correlated contribution $\Gamma_{ij}^c$ can
thus be calculated accurately by exploiting quadrature rules
\begin{equation}\label{eq:density_correlated_part}
\begin{split}
\Gamma_{ij}^c
= \frac{1}{2\pi} \int_{-\infty}^{+\infty} d\omega \, G_{ij}^c(\ii\omega)
= \frac{1}{2\pi} \sum\limits_{k=1}^N \gamma_k {\rm Re}[G_{ij}^c(\ii\omega_k)],
\end{split}
\end{equation}
where $\{\ii\omega_k\}_{k=1}^N$ and $\{\gamma_k\}_{k=1}^N$  are
precalculated imaginary frequency grid points and corresponding weights,
respectively~\cite{KaltakJCTC2014}. Knowing the density matrix, the
particle number is calculated by
\begin{equation}
\label{eq:particle_number}
N_p = {\rm Tr}[\Gamma],
\end{equation}
where the trace ${\rm Tr}$ involves the summation over bands,
$k$-points and spins. The particle number will be taken as an indicator
of the convergence in the self-consistency.

In order to calculate the polarizability $\chi$ and self-energy $\Sigma$
in real space and imaginary time, a Fourier transformation (FT) of $G$
from imaginary frequency to imaginary time is needed. Again, direct FT
of the interacting Green's function $G(\ii\omega)$ is ill-defined.
Therefore, we follow the same strategy that was used when determining  the
density matrix in Eq.~(\ref{eq:split_G_into_two}). Thus, $G(\ii\tau)$
also comprises two parts,
\begin{eqnarray}
\label{eq:G_tau_split_two_parts}
G(\ii\tau)=G^{\rm HF}(\ii\tau)+G^c(\ii\tau).
\end{eqnarray}
In addition, we have used the definitions of the occupied ($\OG$) and
unoccupied ($\UG$) Green's functions as in Ref.~\cite{liu:2016}, which
are evaluated for negative and positive imaginary time, respectively.
With $\OG$ and $\UG$, $G$ can be expressed as
\begin{equation}
\label{eq:single_particle_G}
G(\ii\tau) =\theta(-\tau) \, \OG(\ii\tau) + \theta(\tau) \, \UG(\ii\tau).
\end{equation}
The evaluation of $G^{\rm HF}(\ii\tau)$ is straightforward, since it is
diagonal in the HF canonical-orbital basis,
\begin{eqnarray}
\label{eq:G_tau_HF_occ}
\OG^{\rm HF}_{ij}(\ii\tau) =& \delta_{ij} \,
  \ee^{-(\epsilon_i^{\rm HF}-\mu)\tau} \quad (i,j \in {\rm occ}),  \\
\label{eq:G_tau_HF_unocc}
\UG^{\rm HF}_{ij}(\ii\tau) =& -\delta_{ij} \,
  \ee^{-(\epsilon_i^{\rm HF}-\mu)\tau} \quad (i,j \in {\rm unocc}).
\end{eqnarray}
The correlated $G^c(\ii\tau)$ can be efficiently calculated by inverse
discrete cosine and sine transformations~\cite{KaltakJCTC2014}
\begin{equation}
\begin{split}
\label{eq:G_tau_correlated_occ}
\OG^c_{ij}(\ii\tau_m)
=&
  \sum \limits_{n=1}^{N} \xi_{mn} \cos(\tau_m \omega_n) \,
  \text{Re}\left[G^c_{ij}(\ii \omega_n)\right] \\
-&
  \sum \limits_{n=1}^{N} \zeta_{mn} \sin(\tau_m \omega_n) \,
  \text{Im}\left[G^c_{ij}(\ii \omega_n)\right],
\end{split}
\end{equation}
\begin{equation}
\begin{split}
\label{eq:G_tau_correlated_unocc}
\UG^c_{ij}(\ii\tau_m)
=&
 \sum\limits_{n=1}^{N} \xi_{mn} \cos(\tau_m \omega_n) \,
  \text{Re} \left[G^c_{ij}(\ii \omega_n)\right] \\
+&
\sum\limits_{n=1}^{N}\zeta_{mn} \sin(\tau_m\omega_n) \,
  \text{Im} \left[G^c_{ij}(\ii \omega_n)\right].
 \end{split}
\end{equation}
Here, $\{\ii\tau_m\}_{m=1}^N$ are optimized imaginary time grid points
and the coefficients ${\xi}$ and ${\zeta}$ are precalculated and
stored~\cite{KaltakJCTC2014,liu:2016}.

After the matrices $\OG_{ij}(\ii\tau)$ and $\UG_{ij}(\ii\tau)$ in the HF
canonical-orbital basis have been obtained, they are transformed to the
natural-orbital basis using the unitary matrix $U$ that diagonalizes the interacting
density matrix $\Gamma$ in Eq.~(\ref{eq:definition_density_matrix})
\begin{eqnarray}\label{eq:transform_HF_to_natural}
G_{mn}(\ii\tau) = \sum\limits_{ij}[U^\dag]_{mi} \, G_{ij}(\ii\tau) \, U_{jn}.
\end{eqnarray}
Moreover, the HF canonical orbitals $| \psi^{\rm HF}_{j\kk} \rangle$ are
rotated to the natural orbitals as well,
\begin{eqnarray}\label{eq:transform_HF_to_natural_orbital}
 | \psi_\mk \rangle  =  \sum\limits_{j} U_{jm} \, | \psi^{\rm HF}_{j\kk} \rangle,
\end{eqnarray}
since it is more convenient to evaluate the charge density and the new
HF Hamiltonian $H^{\rm HF}$ in the basis that diagonalizes the interacting density matrix, that is, in the natural orbital basis.

Within the PAW method~\cite{BlochlPAW1994,KressePAW1999},
$G_{mn}(\ii\tau)$ are then transformed from the natural-orbital basis to
the plane-wave (PW) basis by
\begin{eqnarray}
\label{eq:transform_orbital_PW_1}
G^{(1)}_\kk(\sg,\lG', \ii\tau) &=&
  \sum\limits_{m,n}
  \langle \sg |\tilde{\psi}_\mk \rangle \,
  G_{mn}(\ii\tau) \,
  \langle\tilde{\psi}_\nk | \lG' \rangle \\
\label{eq:transform_orbital_PW_2}
G^{(2)}_\kk(\nu,\lG', \ii\tau) &=&
  \sum\limits_{m,n}
  \langle \tilde{p}_\nu |\tilde{\psi}_\mk \rangle \,
  G_{mn}(\ii\tau) \,
  \langle\tilde{\psi}_\nk | \lG' \rangle \\
\label{eq:transform_orbital_PW_3}
G^{(3)}_\kk(\sg,\alpha', \ii\tau) &=&
  \sum\limits_{m,n}
  \langle \sg |\tilde{\psi}_\mk \rangle \,
  G_{mn}(\ii\tau) \,
  \langle\tilde{\psi}_\nk | \tilde{p}_\alpha' \rangle \\
\label{eq:transform_orbital_PW_4}
G^{(4)}_\kk(\nu,\alpha', \ii\tau) &=&
  \sum\limits_{m,n}
  \langle \tilde{p}_\nu |\tilde{\psi}_\mk \rangle \,
  G_{mn}(\ii\tau) \,
  \langle\tilde{\psi}_\nk | \tilde{p}_\alpha' \rangle,
\end{eqnarray}
where $\tilde\psi_\nk$ are pseudo natural orbitals
and $\tilde{p}_\mu$ are projectors, which are dual to the
pseudo partial waves $\tilde\phi_\mu$ within the augmentation
sphere~\cite{BlochlPAW1994,KressePAW1999}.

Knowing $G(\ii\tau)$ in imaginary time, the polarizability
$\chi(\ii\tau)$ is obtained by contraction over $\OG$ and $\UG$ in real space and imaginary time
\begin{equation}
\chi(\ii\tau) = G(\ii\tau) \, G(-\ii\tau)   \quad (\tau > 0),
\end{equation}
which is immediately transformed to $\chi(\ii\omega)$ in imaginary
frequency by the cosine transformations~\cite{KaltakJCTC2014,liu:2016},
where the calculations of the correlated screened interactions $W^c$ are
 conveniently done using the RPA
\begin{equation}
W^c(\ii\omega) = \epsilon^{-1}(\ii\omega) V - V.
\end{equation}
Here $V$ is the bare Coulomb interaction kernel and the inverse of the
dielectric function is calculated by
\begin{equation}
\epsilon^{-1}(\ii\omega)  = 1 + V \chi^{\rm red}(\ii\omega),
\end{equation}
with the reducible polarizability $\chi^{\rm red}$ given by
\begin{equation}
\chi^{\rm red}(\ii\omega) = \left[
  1 - \chi(\ii\omega) V \right]^{-1}\chi(\ii\omega).
\end{equation}
The $W^c(\ii\omega)$ is then transformed to $W^c(\ii\tau)$ by the
inverse of cosine transformations~\cite{KaltakJCTC2014,liu:2016}.
Finally, the new correlated self-energy $\Sigma^c$ is evaluated by the
contraction of $G$ and $W^c$ in real space and imaginary time
\begin{equation}
\Sigma^c(\ii\tau) = -G(\ii\tau) W^c(\ii\tau).
\end{equation}
For the calculations of $\chi(\ii\tau)$ and $\Sigma^c(\ii\tau)$ within
the PAW method, we refer the reader to our previous
publication~\cite{liu:2016}. $\Sigma^c(\ii\tau)$ is then transformed to
$\Sigma^c(\ii\omega)$ in the imaginary frequency domain by the cosine
and sine transformations~\cite{liu:2016}. With the new
$\Sigma^c(\ii\omega)$ and $H^{\rm HF}$, the self-consistency loop of sc$GW$ is closed.
This procedure is repeated until convergence is achieved.

It should be noted that in the second iteration and beyond, the
correlated self-energy $\Sigma^c(\ii\omega)$ is always first evaluated
in the natural-orbital basis and then transformed to the
HF canonical-orbital basis, where an analytic continuation is performed via
a Pad\'{e} fit~\cite{pade1975} to obtain the QP energies and spectral
functions.

%--------------------------------------------------------------------------------
\subsection{Head of the dielectric function}\label{sec:head}
%--------------------------------------------------------------------------------

\begin{figure}
\begin{center}
\includegraphics[width=0.48\textwidth]{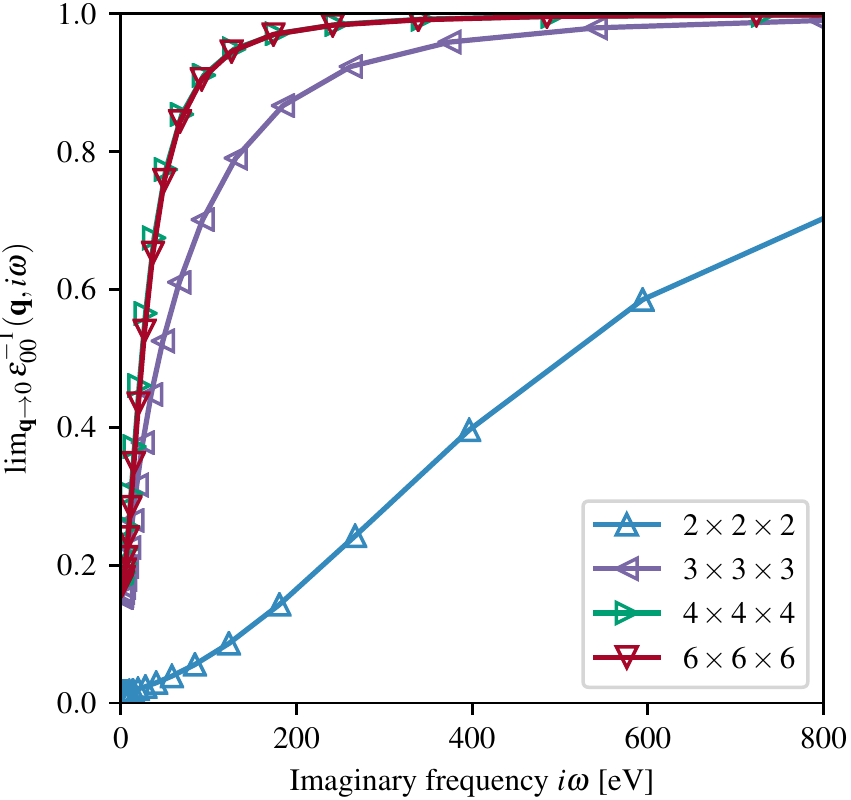}
\end{center}
\caption{
  $k$-point convergence of $\epsilon_{0,0}^{-1}(\mathbf{q}, \ii\omega)$
  in the long-wavelength limit. The data were obtained from sc$GW$
  calculations for diamond.
}
\label{fig:head_convergence}
\end{figure}

In the long-wavelength limit ($\mathbf{q}\rightarrow0$), a
special treatment needs to be done for the head of
$\epsilon_{\lG,\lG'}^{-1}(\qq,\ii\omega)$ (corresponding to $\lG = \lG' = 0$)
due to the singularity of the bare Coulomb interaction.
Within $G_0W_0$, this issue can be tackled
through a Taylor expansion of KS one-electron energies
and orbitals around $\qq = 0$~\cite{Baroni1986,gajdos:2006, harl:2010}.
It can be shown~\cite{Baroni1986,gajdos:2006, harl:2010} that the KS
polarizability $\chi^{\rm KS}_{0,0}(\mathbf{q},\ii\omega)$ at small
$\qq$ for gapped systems has the behavior~\cite{note}
\begin{equation}
\label{eq:chi_expansion}
\chi^{\rm KS}_{0,0}(\qq,\ii\omega) = a \qq^2 + b \qq^4 + \mathcal{O}(\qq^6),
\end{equation}
where $a$ and $b$ are $\qq$-independent constants that can be evaluated
explicitly~\cite{Baroni1986,gajdos:2006, harl:2010}. This leads to the
disappearance of the divergence in $\lim_{\qq \to 0}\epsilon_{0,0}^{-1}(\qq,\ii\omega)$
because of the cancellation of the
$1/\mathbf{q}^2$ terms in the bare Coulomb interaction. However, this
expansion is not possible in subsequent iterations of a sc$GW$
calculation and thus a different solution is needed.

Assuming Eq.~(\ref{eq:chi_expansion}) holds true for the general
polarizability $\chi_{0,0}(\qq,\ii\omega)$, the head
of $\epsilon^{-1}$ has the form
\begin{equation}
\begin{aligned}
\epsilon^{-1}_{00}(\qq, \ii\omega) =
A + B \qq^2 + \mathcal{O}(\qq^4).
\end{aligned}
\end{equation}
The parameters $A$ and $B$ are obtained from a
linear least-square fit on the data from finite $\qq$. The resulting fit
is then extrapolated to $\qq=0$ to estimate the head of
$\epsilon_{\lG,\lG'}^{-1}(\qq,\ii\omega)$ in the long-wavelength limit.
We note that, in principle, $\lim_{\qq \to 0}\epsilon_{0,0}^{-1}(\qq,\ii\omega)$
is a tensor depending on the direction from which $\qq$ approaches zero,
but unfortunately our proposed strategy can only approximately yield the average of diagonal
elements of the tensor.  However, this makes it suitable for the cubic systems
considered in the present work.

In order to check this extrapolation scheme, we compared the extrapolated
$\lim_{\qq \to 0}\epsilon_{0,0}^{-1}(\qq,\ii\omega)$  to the results from $G_0W_0$
calculations, where $\lim_{\qq \to 0}\epsilon_{0,0}^{-1}(\qq,\ii\omega)$ is available
from the above-mentioned $\kk\cdot\mathbf{p}$ perturbation theory.
It was found that the extrapolated $\lim_{\qq \to 0}\epsilon_{0,0}^{-1}(\qq,\ii\omega)$
for diamond using a $6\times6\times6$ $k$-point grid
are almost identical to the ones from  perturbation theory.
This justifies our extrapolation scheme and we can thus apply it to the
sc$GW$ calculations.

The accuracy of the extrapolation scheme also depends on the $k$-point
sampling used, because higher $k$-point densities yield more data points
in the region of small finite $\qq$. As an illustration, Fig.
\ref{fig:head_convergence} shows the convergence of $\lim_{\qq \to
0}\epsilon_{0,0}^{-1}(\mathbf{q}, \ii\omega)$ of diamond with respect to
$k$-point sampling. It can be seen that a $2\times2\times2$ $k$-point
mesh is far from sufficient to obtain the converged head. Results from
$3\times3\times3$ are improved, but still not satisfactory.
Convergence seems to be achieved at $4\times4\times4$ $k$-points
and the results from $6\times6\times6$ $k$-points
are almost unchanged compared to $4\times4\times4$.

We point out here that the inclusion of the head corrections in the
sc$GW$ calculations are important to obtain the precise
QP energies and spectral functions as compared to
calculations without head corrections (see the results in
Sec.~\ref{sec:results}).

\begin{figure*}
\begin{center}
\includegraphics*[width=0.90\textwidth]{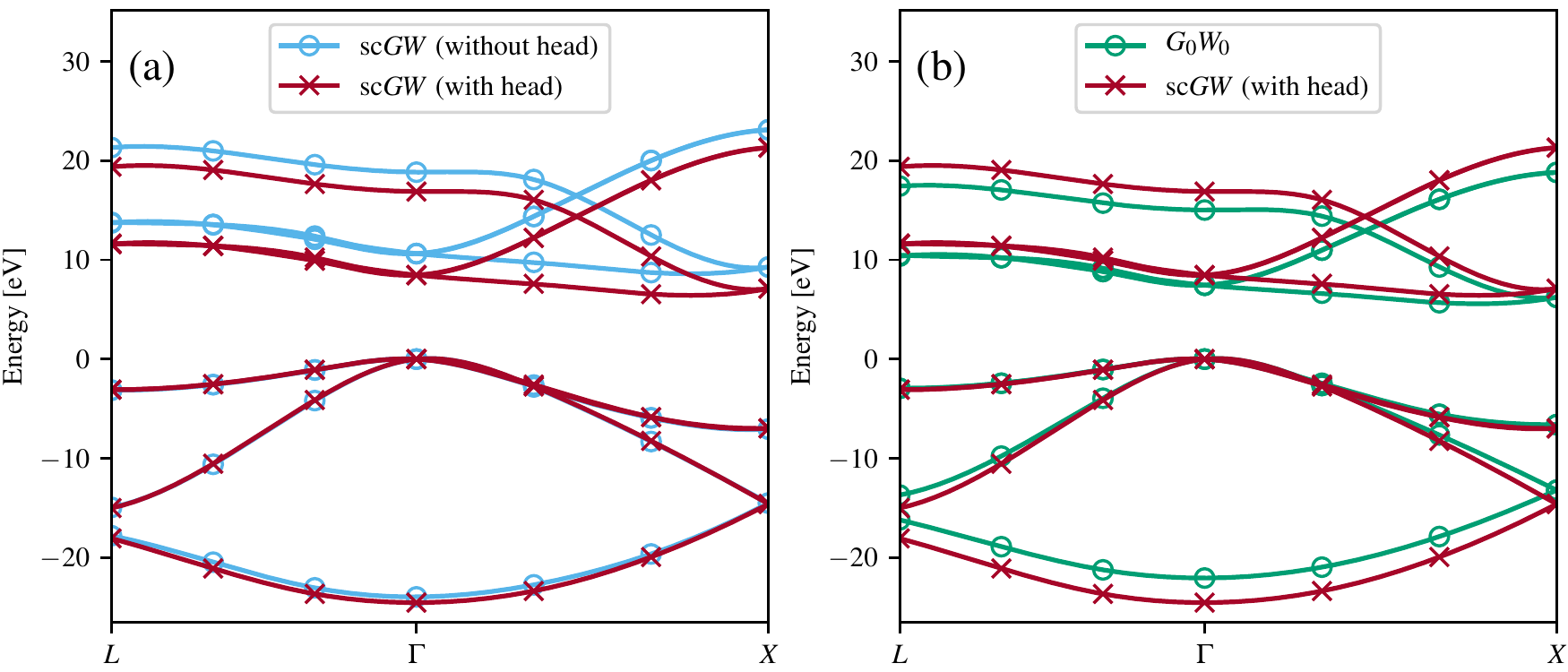}
\end{center}
\caption{
  (a) Comparison of the sc$GW$ QP band structure of diamond with and without
  head corrections in the dielectric function for $6\times 6 \times 6$ $k$-points. (b) sc$GW$ QP band
  structure with head corrections versus $G_0W_0$ QP band structure. The
  data shown here do not include basis-set corrections and $k$-point
  corrections. The VBM at the $\Gamma$ point has been aligned to zero.
  The smooth lines were obtained using a cubic-spline
  interpolation.
}
\label{fig:bandstructure}
\end{figure*}

%--------------------------------------------------------------------------------
\section{Technical Details}\label{sec:details}
%--------------------------------------------------------------------------------

The sc$GW$ method has been implemented in the Vienna \emph{Ab initio}
Simulation Package (VASP)~\cite{KresseVASP1993, KresseVASP1996}. All
calculations were performed using approximately norm-conserving (NC)
$GW$ PAW potentials, the details of which are given in Table I of
Ref.~\cite{klimes:2014}. Table~\ref{table:details} lists all the 15
semiconductors and insulators considered in this work with their
respective crystal structures, lattice constants at low temperature (if
available, otherwise at room temperature) and plane-wave energy cutoffs
of the potentials, which were chosen to be the maximum of all elements in
the considered material. The energy cutoff for the response function was
chosen to be half of the PW cutoff. The number of bands was set to be
the maximum number of PWs compatible to a given PW cutoff energy. To
sample the Brillouin zone, $6\times6\times6$ $k$-point grids centered at
the $\Gamma$ point were used, unless explicitly stated otherwise. The
number of imaginary time/frequency points was set to 24 for
all materials. Five sc$GW$ iterations were performed,
which was found to be sufficient to converge the QP energies to within 0.01 eV. For comparison,
independent QP$GW$ calculations~\cite{shishkin:2007a} were also
performed with 128 real frequency points and a maximum of 6 iterations.

\begin{table}
\caption{
  Crystal structures, lattice constants $a$ and plane-wave energy cutoffs
  $E_{\rm cut}^{\rm pw}$ for all the materials considered.
}
\begin{ruledtabular}
\begin{tabular}{lcrr}
     & Crystal structure & $a$ [\AA] & $E_{\rm cut}^{\rm pw}$ [eV] \\
\hline
BN   & zinc blende       &      3.61 & 700.00 \\
C    & diamond           &      3.56 & 741.69 \\
SiC  & zinc blende       &      4.35 & 741.69 \\
MgO  & rock salt         &      4.21 & 821.52 \\
GaN  & zinc blende       &      4.53 & 801.99 \\
ZnO  & zinc blende       &      4.58 & 802.27 \\
Si   & diamond           &      5.43 & 609.83 \\
AlP  & zinc blende       &      5.46 & 616.62 \\
AlAs & zinc blende       &      5.66 & 613.91 \\
InP  & zinc blende       &      5.86 & 616.62 \\
AlSb & zinc blende       &      6.13 & 571.80 \\
CdS  & zinc blende       &      5.81 & 657.51 \\
ZnS  & zinc blende       &      5.40 & 802.27 \\
GaP  & zinc blende       &      5.45 & 801.99 \\
InSb & zinc blende       &      6.47 & 561.76 \\
\end{tabular}
\end{ruledtabular}
\label{table:details}
\end{table}

In all cases except InSb, standard KS-DFT calculations employing the
Perdew-Burke-Ernzerhof (PBE) functional~\cite{PBE1996,Perdew1997} were used as
starting points for the sc$GW$ and QP$GW$ calculations. For InSb, however, the
hybrid-functional HSE06~\cite{hse:2003,PaierHSE2005} was used instead, because in
this case PBE yields an even qualitatively wrong (negative) band gap. It needs to be mentioned
that the starting point (PBE or HSE06 functional) is only relevant to $G_0W_0$
results due to its first order perturbative nature, whereas for sc$GW$ and QP$GW$
calculations the results are independent of the starting functional.

Since the convergence of the QP energies with respect to the basis set
is slow, we have exploited a basis-set correction
scheme~\cite{klimes:2014, MaggioGW1002017, ergoenenc:2018} using the fact that
the basis-set incompleteness error $E^{\rm QP}(\infty) -E^{\rm QP}(N_{\rm pw})$
decays as $1/N_{\rm pw}$, where $N_{\rm pw}$ is
the number of PWs. Specifically, the PW energy cutoff $E_{\rm cut}^{\rm
pw}$ is increased by a factor of 1.25 and 1.587, leading to an increase
in $N_{\rm pw}$ by a factor of 1.4 and 2.0, respectively. The obtained
results are fitted as a linear function of $1/N_{\rm pw}$ and then
extrapolated to $1/N_{\rm pw} = 0$ to get the final basis-set corrected
QP energies. This was done for both sc$GW$ and QP$GW$ calculations.
Because the sc$GW$ and QP$GW$ calculations are rather demanding and the
basis-set corrections depend only weakly on the number of
$k$-points~\cite{klimes:2014,ergoenenc:2018}, the basis-set corrections
were performed with $3\times3\times 3$ $k$-points. The basis-set
corrected QP energies at $6 \times 6 \times 6$ $k$-points are then
obtained by
\begin{equation}
E_\infty^{6 \times 6 \times 6} =
E_{\rm red}^{6 \times 6 \times 6} +
E_\infty^{3 \times 3 \times 3} -
E_{\rm red}^{3 \times 3 \times 3},
\end{equation}
where $E_{\rm red}^{6 \times 6 \times 6}$  and $E_{\rm red}^{3 \times 3
\times 3}$ are the QP energies calculated using default energy cutoffs
shown in Table~\ref{table:details} for $6 \times 6 \times 6$ and $3
\times 3 \times 3$ $k$-points, respectively, and $E_\infty^{3 \times 3
\times 3}$ is the basis-set corrected QP energy. It was found that the
basis-set corrections are generally small ($<$100 meV), except in a few
cases such as ZnO where they can be as large as a few hundred
meV~\cite{PhysRevLett.105.146401,PhysRevB.83.081101,klimes:2014}.

A similar extrapolation scheme was used to correct the errors introduced by
the $k$-point sampling under the assumption that the $k$-point
set error behaves as $1/N_k$ with $N_k$ being the total number of
$k$-points used. To this end, additional calculations were performed
using a $4 \times 4 \times 4$ $k$-point mesh. The final converged QP
energies including both basis-set and $k$-point set corrections were
obtained by
\begin{equation}
E_\infty^\infty =
E_\infty^{6 \times 6 \times 6} +
E_{\rm red}^\infty -
E_{\rm red}^{6 \times 6 \times 6},
\end{equation}
where the $k$-point corrections $E_{\rm red}^\infty - E_{\rm red}^{6
\times 6 \times 6}$ were calculated with the reduced basis set
corresponding to the default PW cutoff energy, based on the observation
that the $k$-point corrections depend only weakly on the basis
set~\cite{ergoenenc:2018}. Typically, the $k$-point corrections were found
to be on the same order of magnitude as the basis-set corrections.

%--------------------------------------------------------------------------------
\section{Results}\label{sec:results}
%--------------------------------------------------------------------------------

\begin{table}
\caption{
  Comparison of calculated sc$GW$ QP energies of diamond with and without head
  corrections for the dielectric function using $6\times 6 \times 6$ $k$-points. Note that the basis-set and
  $k$-point set corrections are not included here. $\Delta {\rm IP}$  is
  the absolute shift of the VBM at the $\Gamma$ point
  compared to PBE calculations.
  Since the changes in the density and the electrostatic potential
  are small from PBE to sc$GW$, $\Delta {\rm IP}$ is expected to be fairly pseudo/PAW potential independent.
  The other columns represent the relative
  position of the valence band minimum at the $\Gamma$ point ($\Gamma_{\rm
  VBmin}$), the VBM at the $L$ and $X$ points ($L_v$ and $X_v$), and the
  CBM at the $\Gamma$, $L$ and $X$ points
  ($\Gamma_c$, $L_c$ and $X_c$) relative to the VBM at $\Gamma$.
  All values are given in eV.
}
\begin{ruledtabular}
\begin{tabular}{lrrrrrrr}
             & $\Delta {\rm IP}$ & $\Gamma_{\rm VBmin}$ & $\Gamma_{c}$ \
             & $L_{v}$ & $L_{c}$ & $X_{v}$ & $X_{c}$ \\
\hline
Without head    & $-2.47$ & $-23.88$ & 10.64 & $-3.13$ & 13.75 & $-7.07$ & 9.29 \\
With head          & $-1.37$ & $-24.55$ &  8.46 & $-3.08$ & 11.59 & $-6.98$ &  7.10
\end{tabular}
\end{ruledtabular}
\label{table:headcorr}
\end{table}

We present our sc$GW$ results first for the selected case of diamond to
show the effects of the inclusion of the head of the dielectric function
obtained from the extrapolation scheme described in Sec.~\ref{sec:head}
and of the self-consistency on particle number, QP energies, and
spectral functions. Then we extend our discussions to all other
materials.

Table~\ref{table:headcorr} shows a comparison of the calculated sc$GW$
QP energies for diamond at selected $k$-points with and without the head of the
dielectric function. It can be seen that inclusion of the head corrections
changes the QP energies substantially, especially
for deep states (see $\Gamma_{\rm VBmin}$)
and unoccupied states (see $\Gamma_c$, $L_c$ and $X_c$).
Without the head corrections, the band gap
is significantly overestimated as compared to the case with the head corrections.
This is more obviously seen from Fig.~\ref{fig:bandstructure}(a),
where the comparison of the sc$GW$ QP band structures
with and without head corrections are shown.
In addition, it is found that with head corrections, the convergence
of the particle number is faster than without head corrections (not shown).
These findings indicate that the inclusion of the head of the dielectric function
is crucial to obtain accurate and converged QP energies in sc$GW$ calculations
with a reasonable $k$-point set.

\begin{figure}
\begin{center}
\includegraphics*[width=0.45\textwidth]{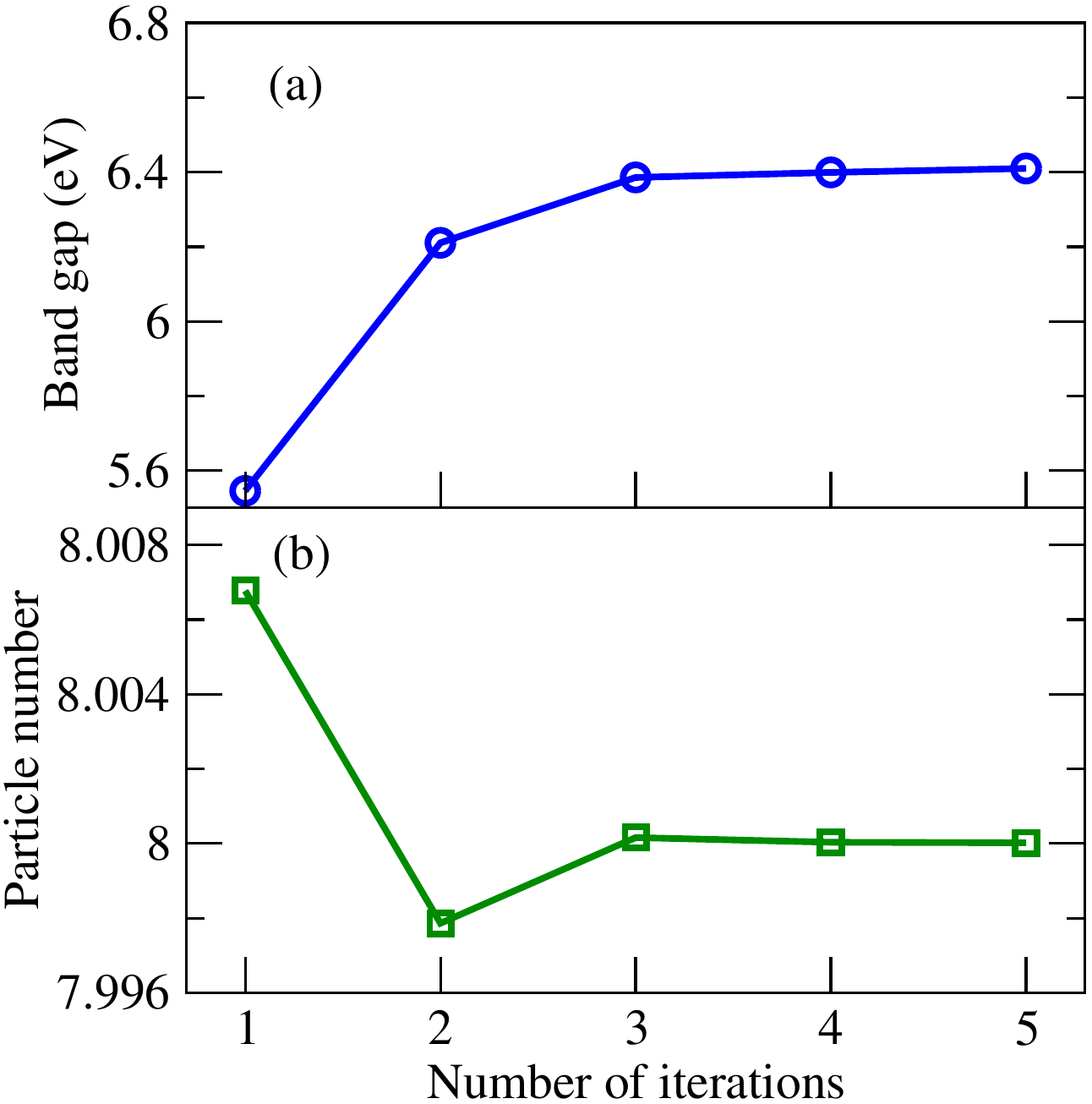}
\end{center}
\caption{
  Calculated (a) band gap and (b) total particle number of diamond from sc$GW$
  calculations as a function of the number of iterations.
  The particle number was obtained from the interacting density matrix [Eq.~(\ref{eq:particle_number})]
  after each iteration.
}
\label{fig:iterations}
\end{figure}

Next, we turn to discuss the effects of self-consistency.
Fig.~\ref{fig:iterations} shows the calculated sc$GW$ band gap and total
particle number of diamond as a function of the iterations. First, it can be seen
that convergence has already been achieved at the fourth iteration.
The converged particle number is calculated to be 8 to an accuracy of
$10^{-4}$, evidencing that sc$GW$ satisfies the conservation of
particle number from a numerical point of view~\cite{Baym1961,Baym1962,Adrian2009}.
As the number of iterations increases, the band gap
increases and finally reaches the converged value of 6.41 eV up to a
precision of 10 meV. We note in passing that our calculated band gap is
about 0.26 eV larger than the sc$GW$ result of A. L.
Kutepov~\cite{kutepov:2017}. The discrepancy might arise from different
implementation and setup details, such as potentials, basis sets and
$k$-point grid used. Also, it is not clear how the head of the
dielectric function was dealt with in Ref.~\cite{kutepov:2017}. Second,
one can see that the sc$GW$ band gap is significantly enlarged
compared to $G_0W_0$. This is more clearly seen from the QP band
structure comparison between sc$GW$ and $G_0W_0$ in
Fig.~\ref{fig:bandstructure}(b). This is expected because our full
sc$GW$ does not take into account vertex corrections. Indeed,
inclusion of vertex corrections in sc$GW$ will reduce the gap towards
the experimental value~\cite{kutepov:2017,grueneis:2014}, but this is
beyond the scope of the present work.

\begin{figure}
\begin{center}
\includegraphics*[width=0.48\textwidth]{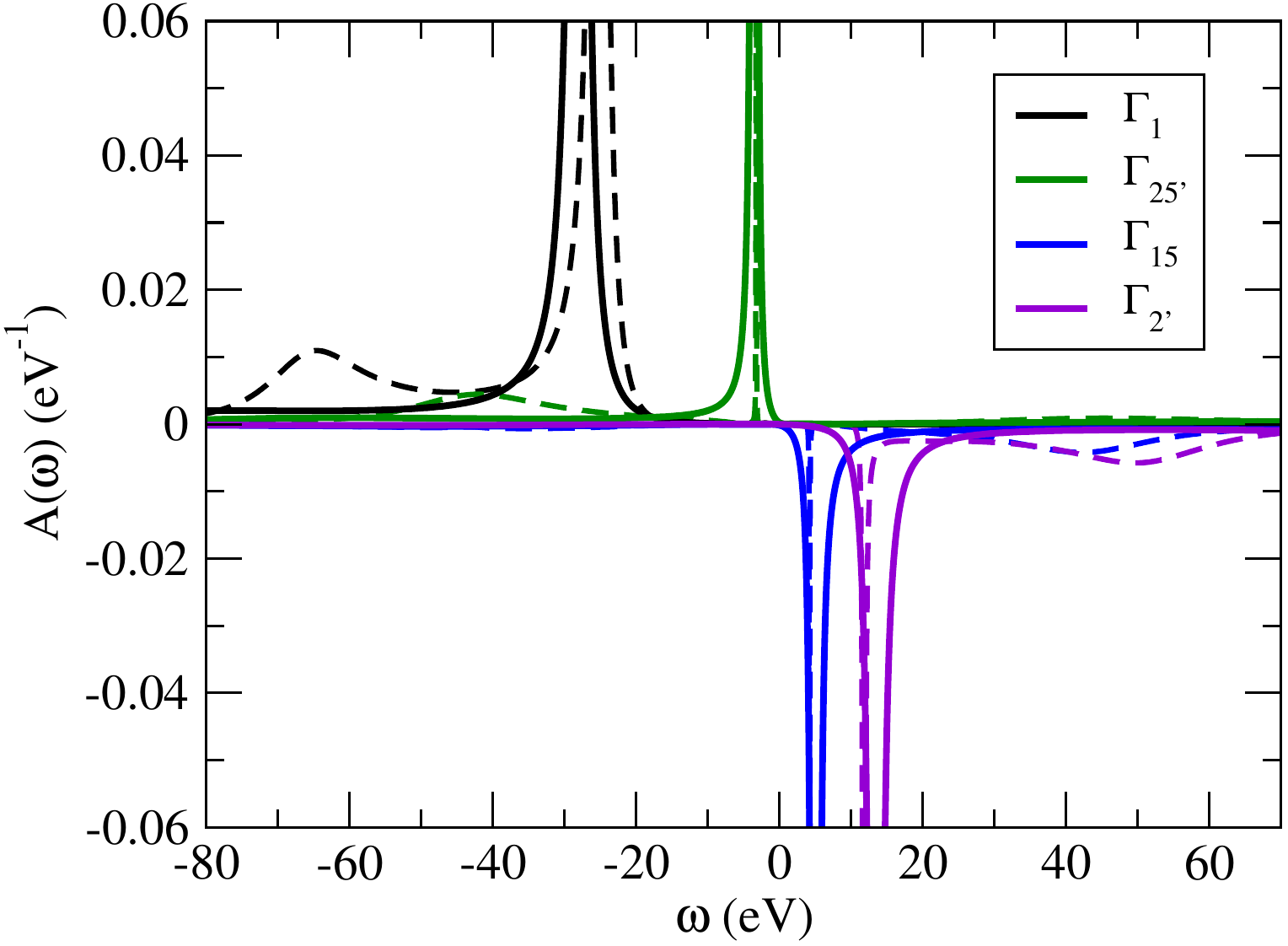}
\end{center}
\caption{
  Comparison of sc$GW$ (solid lines) and $G_0W_0$ (dashed lines) spectral
  functions of diamond for selected bands at $\Gamma$. Note that the
  signs of the spectral functions for unoccupied states such as
  $\Gamma_{15}$ and $\Gamma_{2'}$ are intentionally reversed for
  clarity.
}
\label{fig:C_spectral}
\end{figure}

\begin{table}
\caption{
The renormalization factor $Z$ of the QP peaks of diamond for selected bands at $\Gamma$
predicted by $G_0W_0$ and sc$GW$.
}
\begin{ruledtabular}
\begin{tabular}{lcccc}
             & $\Gamma_1$ & $\Gamma_{25'}$ & $\Gamma_{15}$  & $\Gamma_{2'}$ \\
\hline
$G_0W_0$    &  0.753  &0.827 & 0.828 & 0.771 \\
sc$GW$         & 0.885   &0.884 & 0.897 & 0.903
\end{tabular}
\end{ruledtabular}
\label{table:Z}
\end{table}

Fig.~\ref{fig:C_spectral} shows the sc$GW$ spectral functions of diamond for
selected bands at the $\Gamma$ point along with $G_0W_0$ spectral functions.
One can observe that, compared to $G_0W_0$, the QP peaks of the spectral
functions from sc$GW$ at the $\Gamma$ point are broadened and
dramatically shifted up and down for unoccupied and occupied bands,
respectively, leading to a significant enhancement of the direct band
gap. This is consistent with the QP band structures shown in
Fig.~\ref{fig:bandstructure}(b).
In addition, we find that the satellite
structures (plasmonic polarons) appearing far below or above the QP peaks in
$G_0W_0$ are washed out in sc$GW$, in accordance with the increased
renormalization factor $Z$ of QP peaks predicted by sc$GW$ (see Table~\ref{table:Z}).
It is known that these satellites
are to some extent artifacts of the $G_0W_0$ approximation, although similar
features are observed in X-ray photo-emission experiments~\cite{Guzzo_CE2011,Caruso2015}.
In the experiment the features are interpreted as replicas of the QP peak shifted by the typical
plasmon frequency. The complete absence of these features
in the sc$GW$ is troublesome and clearly
suggests that one needs to go beyond the $GW$ approximation,
for example, by the cumulant expansion of the Green's function
~\cite{aryasetiawan:1996, Guzzo_CE2011, kas:2014, Caruso2015, gumhalter:2016}
or by including an approximate vertex.

\begin{table}
\caption{
  QP energies and fundamental band gaps ($\Delta_{\rm gap}$) (in eV)
  from sc$GW$ calculations. The meaning of each column is the same as in
  Table~\ref{table:headcorr}. Note that for InSb, $\Delta {\rm IP}$ is
  calculated with respect to HSE06 calculations.
}
\begin{ruledtabular}
\footnotesize
\begin{tabular}{lrrrrrrrr}
 & $\Delta \mathrm{IP}$ & $\Gamma_\mathrm{VBmin}$ & $\Gamma_{c}$ & $L_{v}$ & $L_{c}$ & $X_{v}$ & $X_{c}$ & $\Delta_\mathrm{gap}$ \\
\hline
BN           &  $-2.02$ & $-23.42$ &  12.86 &  $-2.07$ &  13.81 &  $-5.27$ &   7.67 &   7.67 \\
SiC          &  $-1.28$ & $-17.96$ &   8.69 &  $-1.18$ &   7.85 &  $-3.56$ &   3.29 &   3.29 \\
C            &  $-1.37$ & $-24.76$ &   8.33 &  $-3.06$ &  11.47 &  $-6.95$ &   6.97 &   6.41 \\
Si           &  $-1.09$ & $-15.58$ &   4.71 &  $-1.43$ &   3.34 &  $-3.48$ &   2.29 &   2.18 \\
AlP          &  $-1.10$ & $-13.03$ &   5.01 &  $-0.82$ &   4.52 &  $-2.26$ &   3.20 &   3.20 \\
AlAs         &  $-1.09$ & $-13.37$ &   3.73 &  $-0.87$ &   3.75 &  $-2.31$ &   2.98 &   2.98 \\
AlSb         &  $-0.93$ & $-12.74$ &   3.18 &  $-1.02$ &   2.78 &  $-2.47$ &   2.61 &   2.61 \\
InP          &  $-0.92$ & $-13.52$ &   1.97 &  $-1.04$ &   2.89 &  $-2.50$ &   3.04 &   1.97 \\
InSb         &  $-0.50$ & $-11.46$ &   0.79 &  $-1.00$ &   1.40 &  $-2.26$ &   2.13 &   0.79 \\
GaN          &  $-1.55$ & $-18.92$ &   3.94 &  $-1.01$ &   7.19 &  $-2.83$ &   5.57 &   3.94 \\
GaP          &  $-0.83$ & $-14.12$ &   3.17 &  $-1.11$ &   2.96 &  $-2.82$ &   2.77 &   2.77 \\
ZnO          &  $-3.04$ & $-20.06$ &   4.92 &  $-0.82$ &  10.40 &  $-2.23$ &   9.52 &   4.92 \\
ZnS          &  $-1.72$ & $-14.43$ &   4.68 &  $-0.86$ &   6.02 &  $-2.24$ &   5.70 &   4.68 \\
CdS          &  $-1.60$ & $-14.60$ &   3.46 &  $-0.84$ &   5.65 &  $-2.09$ &   5.88 &   3.46 \\
MgO          &  $-3.27$ & $-20.08$ &   9.53 &  $-0.80$ &  12.83 &  $-1.65$ &  13.80 &   9.53
\end{tabular}
\end{ruledtabular}
\label{table:scgw}
\end{table}

\begin{table}
\caption{Same as Table~\ref{table:scgw}, but for $G_0W_0$ results.}
\begin{ruledtabular}
\footnotesize
\begin{tabular}{lrrrrrrrr}
 & $\Delta \mathrm{IP}$ & $\Gamma_\mathrm{VBmin}$ & $\Gamma_{c}$ & $L_{v}$ & $L_{c}$ & $X_{v}$ & $X_{c}$ & $\Delta_\mathrm{gap}$ \\
\hline
BN           &  $-1.51$ & $-20.89$ &  11.33 &  $-2.07$ &  12.29 &  $-5.17$ &   6.39 &   6.39 \\
SiC          &  $-1.03$ & $-15.51$ &   7.35 &  $-1.09$ &   6.63 &  $-3.29$ &   2.42 &   2.42 \\
C            &  $-1.22$ & $-21.98$ &   7.44 &  $-2.94$ &  10.39 &  $-6.59$ &   6.24 &   5.69 \\
Si           &  $-0.71$ & $-12.00$ &   3.24 &  $-1.21$ &   2.10 &  $-2.85$ &   1.26 &   1.15 \\
AlP          &  $-0.94$ & $-11.28$ &   4.20 &  $-0.78$ &   3.77 &  $-2.13$ &   2.47 &   2.47 \\
AlAs         &  $-1.01$ & $-11.71$ &   2.97 &  $-0.83$ &   3.07 &  $-2.17$ &   2.30 &   2.30 \\
AlSb         &  $-0.84$ & $-10.64$ &   2.38 &  $-0.91$ &   2.06 &  $-2.21$ &   1.86 &   1.86 \\
InP          &  $-0.79$ & $-11.15$ &   1.26 &  $-0.98$ &   2.10 &  $-2.39$ &   2.32 &   1.26 \\
InSb         &  $-0.58$ & $-10.88$ &   0.57 &  $-1.19$ &   1.23 &  $-2.37$ &   1.98 &   0.57 \\
GaN          &  $-1.19$ & $-15.31$ &   2.87 &  $-0.98$ &   5.95 &  $-2.72$ &   4.54 &   2.87 \\
GaP          &  $-0.85$ & $-12.28$ &   2.62 &  $-1.14$ &   2.45 &  $-2.72$ &   2.31 &   2.31 \\
ZnO          &  $-1.77$ & $-18.19$ &   2.55 &  $-0.83$ &   7.55 &  $-2.20$ &   7.08 &   2.55 \\
ZnS          &  $-1.31$ & $-11.96$ &   3.43 &  $-0.85$ &   4.77 &  $-2.20$ &   4.66 &   3.43 \\
CdS          &  $-1.12$ & $-11.40$ &   2.16 &  $-0.78$ &   4.22 &  $-2.01$ &   4.59 &   2.16 \\
MgO          &  $-2.10$ & $-17.82$ &   7.49 &  $-0.73$ &  10.76 &  $-1.48$ &  11.78 &   7.49
\end{tabular}
\end{ruledtabular}
\label{table:g0w0}
\end{table}

\begin{table}
\caption{Same as Table~\ref{table:scgw}, but for QP$GW$ results.}
\begin{ruledtabular}
\footnotesize
\begin{tabular}{lrrrrrrrr}
 & $\Delta \mathrm{IP}$ & $\Gamma_\mathrm{VBmin}$ & $\Gamma_{c}$ & $L_{v}$ & $L_{c}$ & $X_{v}$ & $X_{c}$ & $\Delta_\mathrm{gap}$ \\
\hline
BN           &  $-2.25$ & $-21.61$ &  12.55 &  $-2.33$ &  13.48 &  $-5.34$ &   7.50 &   7.50 \\
SiC          &  $-1.35$ & $-16.22$ &   7.82 &  $-1.19$ &   7.14 &  $-3.43$ &   2.88 &   2.88 \\
C            &  $-1.81$ & $-22.73$ &   8.03 &  $-3.08$ &  11.20 &  $-6.71$ &   6.97 &   6.43 \\
Si           &  $-1.01$ & $-12.15$ &   3.65 &  $-1.23$ &   2.47 &  $-2.93$ &   1.60 &   1.49 \\
AlP          &  $-1.28$ & $-11.80$ &   4.74 &  $-0.81$ &   4.29 &  $-2.23$ &   2.94 &   2.94 \\
AlAs         &  $-1.43$ & $-12.23$ &   3.58 &  $-0.85$ &   3.62 &  $-2.25$ &   2.84 &   2.84 \\
AlSb         &  $-1.13$ & $-11.02$ &   2.78 &  $-0.94$ &   2.43 &  $-2.28$ &   2.22 &   2.22 \\
InP          &  $-1.13$ & $-11.73$ &   1.64 &  $-1.01$ &   2.53 &  $-2.43$ &   2.71 &   1.64 \\
InSb         &  $-0.77$ & $-10.98$ &   0.61 &  $-1.05$ &   1.31 &  $-2.36$ &   2.05 &   0.61 \\
GaN          &  $-1.74$ & $-16.38$ &   3.78 &  $-1.02$ &   6.91 &  $-2.81$ &   5.39 &   3.78 \\
GaP          &  $-1.12$ & $-12.78$ &   3.05 &  $-1.18$ &   2.86 &  $-2.82$ &   2.67 &   2.67 \\
ZnO          &  $-2.85$ & $-19.02$ &   4.29 &  $-0.83$ &   9.46 &  $-2.29$ &   8.86 &   4.29 \\
ZnS          &  $-1.80$ & $-13.32$ &   4.27 &  $-0.89$ &   5.66 &  $-2.30$ &   5.42 &   4.27 \\
CdS          &  $-1.65$ & $-12.61$ &   2.89 &  $-0.81$ &   4.98 &  $-2.01$ &   5.31 &   2.89 \\
MgO          &  $-3.42$ & $-18.32$ &   9.58 &  $-0.52$ &  12.71 &  $-1.36$ &  13.75 &   9.58
\end{tabular}
\end{ruledtabular}
\label{table:qpgw}
\end{table}

Having validated our sc$GW$ implementation for the selected case of diamond,
we now extend our discussion to all other considered materials. The
calculated sc$GW$ QP energies are compiled in Table~\ref{table:scgw}.
For comparison, the QP energies obtained from $G_0W_0$ and QP$GW$
calculations are also given in Table~\ref{table:g0w0} and
Table~\ref{table:qpgw}, respectively.
We first note that our $G_0W_0$ QP energies agree very well with
the results in Ref.~\cite{klimes:2014} with deviations less than 50 meV
for all the materials considered except for ZnO and InSb.
The deviations arise from the different implementations and setups.
Specifically, our $G_0W_0$ QP energies were obtained from the analytic continuation
of the self-energy in imaginary frequency, whereas the results of Ref.~\cite{klimes:2014}
were computed from the self-energy evaluated along the real frequency axis.
Both methods can result in errors: the analytic continuation is known
to be ill-conditioned, although band gaps are usually very accurate.
On the other hand, calculations along the real frequency axis are prone to discretization errors.  In addition,
to calculate the derivative of the cell-periodic part of the KS orbitals
with respect to $\kk$, $| \nabla_\kk u_\nk \rangle$,
which is needed to deal with the long-wavelength limit of the dielectric function~\cite{gajdos:2006},
the finite difference method by the perturbation expansion after
discretization (PEAD) method~\cite{nunes:2001} was used in
the present work, while in Ref.~\cite{klimes:2014} $| \nabla_\kk u_\nk \rangle$
was obtained by linear response theory~\cite{gajdos:2006}.
Although, both methods strictly converge to the same values
when the number of $k$-points is sufficiently large,
the dielectric function
coverges from below  and above for the PEAD and linear response, respectively.
This means that the band gaps converge from above and below for the PEAD and linear response, respectively.
This explains the larger band gap of ZnO in our $G_0W_0$ calculations as compared
to Ref.~\cite{klimes:2014}.  For the
small $6\times6\times6$ $k$-mesh  used here,  the PEAD method is found to be
more accurate, since the calculated dielectric functions
are already very close to the converged values.
The larger deviation for InSb is understood, because our $G_0W_0$
calculations for InSb were done on top of the HSE06 functional instead of
the PBE calculations used in Ref.~\cite{klimes:2014}.

\begin{figure}
\begin{center}
\includegraphics*[width=0.49\textwidth]{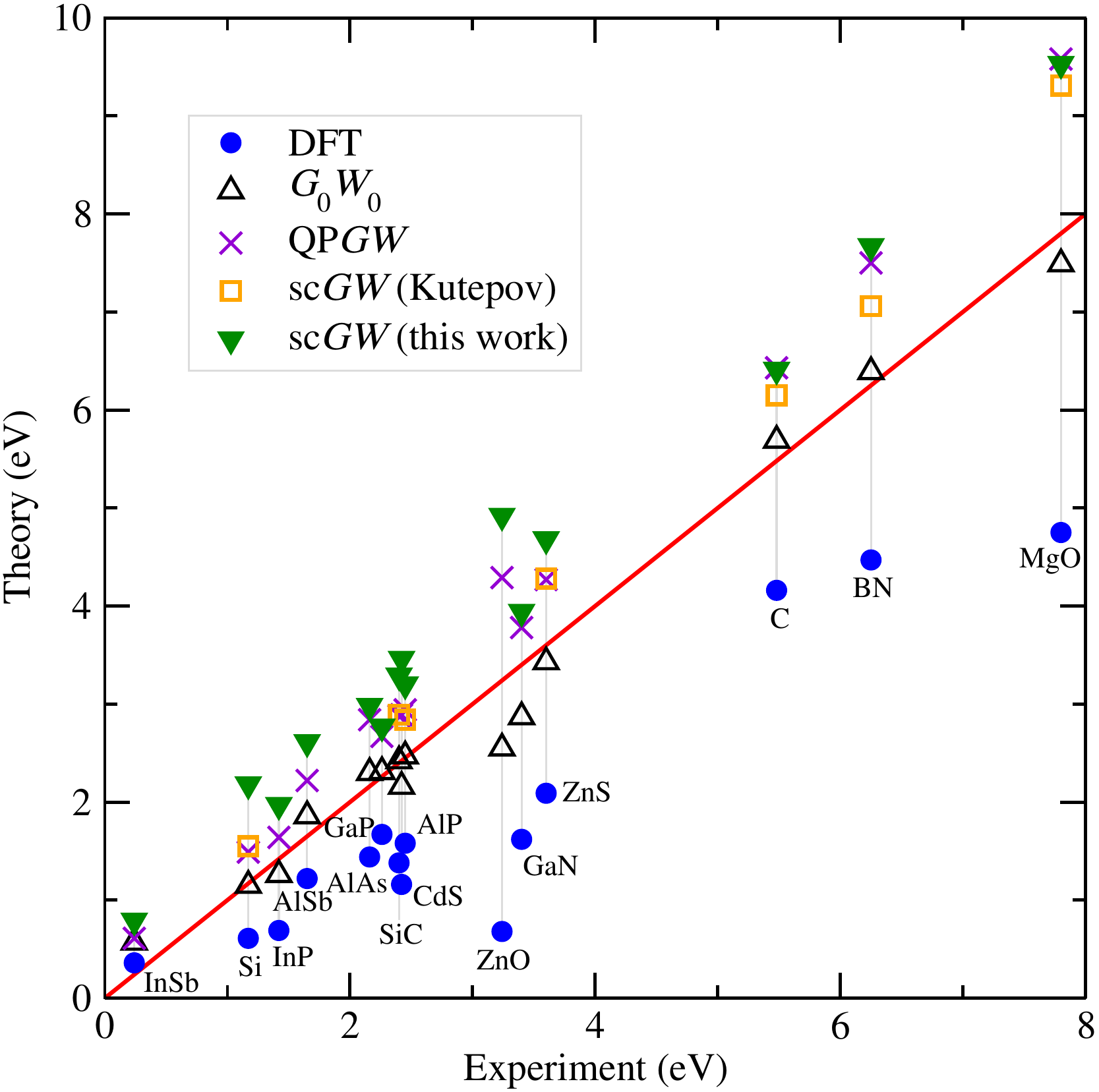}
\end{center}
\caption{
  Comparison of band gaps obtained from different theoretical methods and
  experiments for the whole set of materials considered here. The sc$GW$ results
  of A. L. Kutepov are taken from Ref.~\cite{kutepov:2017}. Experimental
  values are taken from Ref.~\cite{grueneis:2014} and references
  therein. Note that the HSE06 calculated band gap is shown for InSb.
}
\label{fig:bandgaps}
\end{figure}

We now turn to the sc$GW$ and QP$GW$ results.
sc$GW$ and QP$GW$ raise the unoccupied states, but lower the occupied states, in particular
the deep states (see $\Gamma_{\rm VBmin}$ in the tables for instance),
with the shifts being more apparent in sc$GW$.
Fig.~\ref{fig:bandgaps} furthermore shows the band gaps calculated from various theoretical methods
at different levels against experimental results. In addition, the
sc$GW$ results for a subset of materials from A. L.
Kutepov~\cite{kutepov:2017} are also shown for comparison. One can see
that, as expected, PBE underestimates the band gaps due to
neglected integer discontinuity. Inclusion of the non-local dynamical
self-energy in the $G_0W_0$ approximation improves the band gaps towards
the experimental values. Nevertheless, from a fundamental point, $G_0W_0$ is always
somewhat unsatisfactory, since the good
agreement arises almost certainly from the aforementioned cancellation of
errors due to the lack of self-consistency and vertex corrections.
Introducing self-consistency only, however, deteriorates the results,
leading to a significant increase of the sc$GW$ and QP$GW$ band
gaps as compared to $G_0W_0$ and experiment. The overestimation is generally larger in
sc$GW$ than QP$GW$, which is due to the smaller values in the dielectric functions predicted
by sc$GW$. This shows that the RPA is not sufficiently accurate when used on top of sc$GW$
or QP$GW$ calculations. This
was to be expected, since Hedin's equations clearly imply that the interaction kernel must be set to the functional
derivative of the self-energy (here $\Sigma=GW$) with respect to the Green's function $G$.
Neglecting the variation of $W$ with respect to $G$, this implies that even
at the simplest level of theory, one needs to include a screened exchange interaction
$W$ via the Bethe-Salpeter equation in the irreducible polarizability.

Our calculated sc$GW$ band gaps are, in general,
consistent with the results of A. L.  Kutepov~\cite{kutepov:2017} for
the given materials, but are a bit larger, typically by up to 0.4 eV. The
possible reasons for the deviations have already been discussed before.
We want to reiterate here that our sc$GW$ band gaps have been
corrected for the finite basis sets errors as well as errors introduced by
the finite $k$-point sampling.  Also, the singularity problem
associated with the long-wavelength limit of the dielectric function has
been carefully dealt with by the extrapolation scheme.

%--------------------------------------------------------------------------------
\section{Conclusions}\label{sec:conlcusions}
%--------------------------------------------------------------------------------

To summarize, we have presented converged QP energies for
15 semiconductors and insulators from a fully self-consistent $GW$
implementation within the PAW method. Converged band gaps have  been obtained by
including both finite basis-set corrections and $k$-point corrections, as well as
using accurate (norm-conserving) $GW$ PAW potentials. Furthermore,
a simple extrapolation scheme has been used to determine the dielectric
matrix  in the long-wavelength limit.  All
implementation details were given and particular emphasis was put on the
extrapolation scheme.  Our implementation was tested, first by
investigating  the selected case of diamond and then for the entire
set of compounds. The calculated sc$GW$ band gaps were
compared to $G_0W_0$, QP$GW$ and, where available, sc$GW$ band gaps
as well as experimental results. It was found that for the sc$GW$
calculations the inclusion of the head corrections in the dielectric
function is important to obtain reasonably fast convergence of the QP energies.

From a physical point of view, our results can be summarized as follows.
The sc$GW$ method yields mostly unsatisfactory results compared to experiment.
Notably, the band gaps are significantly overestimated compared to experiment,
and plasmonic satellites are entirely missing in the spectral function (see Fig.~\ref{fig:C_spectral}).
As we have explained in the previous section, this is related to the
absence of vertex corrections, which Hedin's equations dictate  to be the
derivative of the $GW$ self-energy with respect to the external potential.
As already discussed by Kutepov~\cite{kutepov:2016,kutepov:2017}, including a
consistent vertex is  a formidable task: even in the simplest approximation, one would need
to include a vertex $W$ via the  Bethe-Salpether equation~\cite{maggio:2017a}, which increases
the compute cost by at least an order of magnitude. Worse, if one would continue
self-consistency, the derivative of the self-energy--- then $\Sigma=GW\Gamma$
(using a very simple $\Gamma$)  ---with respect to the Green's function $G$ would create even more diagrams,
increasing the complexity of the vertex further. So one is necessarily faced with
the dilemma at which level of theory one terminates the cycle. This choice will
be mandated by the computational requirements and the implementational complexity.
Without question, sc$GW$ is unsatisfactory. Nevertheless,  our calculations
establish accurate reference values for sc$GW$, upon which one can now try to improve,
for instance, by including the simplest possible vertex $W$.

\begin{acknowledgments}
This work was supported by the FWF within the SFB ViCoM (Grant No. F 41).
Supercomputing time on the Vienna Scientific cluster (VSC) is gratefully acknowledged.
JK was supported by the European Union's Horizon 2020 research and
innovation programme under the Marie Sklodowska-Curie grant agreement No.~658705.
\end{acknowledgments}

\bibliographystyle{apsrev4-1}
\bibliography{references}

%merlin.mbs apsrev4-1.bst 2010-07-25 4.21a (PWD, AO, DPC) hacked
%Control: key (0)
%Control: author (72) initials jnrlst
%Control: editor formatted (1) identically to author
%Control: production of article title (-1) disabled
%Control: page (0) single
%Control: year (1) truncated
%Control: production of eprint (0) enabled
\begin{thebibliography}{61}%
\makeatletter
\providecommand \@ifxundefined [1]{%
 \@ifx{#1\undefined}
}%
\providecommand \@ifnum [1]{%
 \ifnum #1\expandafter \@firstoftwo
 \else \expandafter \@secondoftwo
 \fi
}%
\providecommand \@ifx [1]{%
 \ifx #1\expandafter \@firstoftwo
 \else \expandafter \@secondoftwo
 \fi
}%
\providecommand \natexlab [1]{#1}%
\providecommand \enquote  [1]{``#1''}%
\providecommand \bibnamefont  [1]{#1}%
\providecommand \bibfnamefont [1]{#1}%
\providecommand \citenamefont [1]{#1}%
\providecommand \href@noop [0]{\@secondoftwo}%
\providecommand \href [0]{\begingroup \@sanitize@url \@href}%
\providecommand \@href[1]{\@@startlink{#1}\@@href}%
\providecommand \@@href[1]{\endgroup#1\@@endlink}%
\providecommand \@sanitize@url [0]{\catcode `\\12\catcode `\$12\catcode
  `\&12\catcode `\#12\catcode `\^12\catcode `\_12\catcode `\%12\relax}%
\providecommand \@@startlink[1]{}%
\providecommand \@@endlink[0]{}%
\providecommand \url  [0]{\begingroup\@sanitize@url \@url }%
\providecommand \@url [1]{\endgroup\@href {#1}{\urlprefix }}%
\providecommand \urlprefix  [0]{URL }%
\providecommand \Eprint [0]{\href }%
\providecommand \doibase [0]{http://dx.doi.org/}%
\providecommand \selectlanguage [0]{\@gobble}%
\providecommand \bibinfo  [0]{\@secondoftwo}%
\providecommand \bibfield  [0]{\@secondoftwo}%
\providecommand \translation [1]{[#1]}%
\providecommand \BibitemOpen [0]{}%
\providecommand \bibitemStop [0]{}%
\providecommand \bibitemNoStop [0]{.\EOS\space}%
\providecommand \EOS [0]{\spacefactor3000\relax}%
\providecommand \BibitemShut  [1]{\csname bibitem#1\endcsname}%
\let\auto@bib@innerbib\@empty
%</preamble>
\bibitem [{\citenamefont {Hedin}(1965)}]{Hedin1965}%
  \BibitemOpen
  \bibfield  {author} {\bibinfo {author} {\bibfnamefont {L.}~\bibnamefont
  {Hedin}},\ }\href {\doibase 10.1103/PhysRev.139.A796} {\bibfield  {journal}
  {\bibinfo  {journal} {Phys. Rev.}\ }\textbf {\bibinfo {volume} {139}},\
  \bibinfo {pages} {A796} (\bibinfo {year} {1965})}\BibitemShut {NoStop}%
\bibitem [{\citenamefont {Hedin}\ and\ \citenamefont
  {Lundqvist}(1969)}]{Hedin1969}%
  \BibitemOpen
  \bibfield  {author} {\bibinfo {author} {\bibfnamefont {L.}~\bibnamefont
  {Hedin}}\ and\ \bibinfo {author} {\bibfnamefont {S.}~\bibnamefont
  {Lundqvist}},\ }\href@noop {} {\emph {\bibinfo {title} {Solid State
  Physics}}}\ (\bibinfo  {publisher} {Academic Press},\ \bibinfo {address} {New
  York},\ \bibinfo {year} {1969})\BibitemShut {NoStop}%
\bibitem [{\citenamefont {Strinati}\ \emph {et~al.}(1980)\citenamefont
  {Strinati}, \citenamefont {Mattausch},\ and\ \citenamefont
  {Hanke}}]{Strinati1980}%
  \BibitemOpen
  \bibfield  {author} {\bibinfo {author} {\bibfnamefont {G.}~\bibnamefont
  {Strinati}}, \bibinfo {author} {\bibfnamefont {H.~J.}\ \bibnamefont
  {Mattausch}}, \ and\ \bibinfo {author} {\bibfnamefont {W.}~\bibnamefont
  {Hanke}},\ }\href {\doibase 10.1103/PhysRevLett.45.290} {\bibfield  {journal}
  {\bibinfo  {journal} {Phys. Rev. Lett.}\ }\textbf {\bibinfo {volume} {45}},\
  \bibinfo {pages} {290} (\bibinfo {year} {1980})}\BibitemShut {NoStop}%
\bibitem [{\citenamefont {Strinati}\ \emph {et~al.}(1982)\citenamefont
  {Strinati}, \citenamefont {Mattausch},\ and\ \citenamefont
  {Hanke}}]{Strinati1982}%
  \BibitemOpen
  \bibfield  {author} {\bibinfo {author} {\bibfnamefont {G.}~\bibnamefont
  {Strinati}}, \bibinfo {author} {\bibfnamefont {H.~J.}\ \bibnamefont
  {Mattausch}}, \ and\ \bibinfo {author} {\bibfnamefont {W.}~\bibnamefont
  {Hanke}},\ }\href {\doibase 10.1103/PhysRevB.25.2867} {\bibfield  {journal}
  {\bibinfo  {journal} {Phys. Rev. B}\ }\textbf {\bibinfo {volume} {25}},\
  \bibinfo {pages} {2867} (\bibinfo {year} {1982})}\BibitemShut {NoStop}%
\bibitem [{\citenamefont {Hybertsen}\ and\ \citenamefont
  {Louie}(1985)}]{Louie1985}%
  \BibitemOpen
  \bibfield  {author} {\bibinfo {author} {\bibfnamefont {M.~S.}\ \bibnamefont
  {Hybertsen}}\ and\ \bibinfo {author} {\bibfnamefont {S.~G.}\ \bibnamefont
  {Louie}},\ }\href {\doibase 10.1103/PhysRevLett.55.1418} {\bibfield
  {journal} {\bibinfo  {journal} {Phys. Rev. Lett.}\ }\textbf {\bibinfo
  {volume} {55}},\ \bibinfo {pages} {1418} (\bibinfo {year}
  {1985})}\BibitemShut {NoStop}%
\bibitem [{\citenamefont {Hybertsen}\ and\ \citenamefont
  {Louie}(1986)}]{Louie1986}%
  \BibitemOpen
  \bibfield  {author} {\bibinfo {author} {\bibfnamefont {M.~S.}\ \bibnamefont
  {Hybertsen}}\ and\ \bibinfo {author} {\bibfnamefont {S.~G.}\ \bibnamefont
  {Louie}},\ }\href {\doibase 10.1103/PhysRevB.34.5390} {\bibfield  {journal}
  {\bibinfo  {journal} {Phys. Rev. B}\ }\textbf {\bibinfo {volume} {34}},\
  \bibinfo {pages} {5390} (\bibinfo {year} {1986})}\BibitemShut {NoStop}%
\bibitem [{\citenamefont {Aryasetiawan}\ \emph {et~al.}(1996)\citenamefont
  {Aryasetiawan}, \citenamefont {Hedin},\ and\ \citenamefont
  {Karlsson}}]{aryasetiawan:1996}%
  \BibitemOpen
  \bibfield  {author} {\bibinfo {author} {\bibfnamefont {F.}~\bibnamefont
  {Aryasetiawan}}, \bibinfo {author} {\bibfnamefont {L.}~\bibnamefont {Hedin}},
  \ and\ \bibinfo {author} {\bibfnamefont {K.}~\bibnamefont {Karlsson}},\
  }\href {\doibase 10.1103/PhysRevLett.77.2268} {\bibfield  {journal} {\bibinfo
   {journal} {Phys. Rev. Lett.}\ }\textbf {\bibinfo {volume} {77}},\ \bibinfo
  {pages} {2268} (\bibinfo {year} {1996})}\BibitemShut {NoStop}%
\bibitem [{\citenamefont {Guzzo}\ \emph {et~al.}(2011)\citenamefont {Guzzo},
  \citenamefont {Lani}, \citenamefont {Sottile}, \citenamefont {Romaniello},
  \citenamefont {Gatti}, \citenamefont {Kas}, \citenamefont {Rehr},
  \citenamefont {Silly}, \citenamefont {Sirotti},\ and\ \citenamefont
  {Reining}}]{Guzzo_CE2011}%
  \BibitemOpen
  \bibfield  {author} {\bibinfo {author} {\bibfnamefont {M.}~\bibnamefont
  {Guzzo}}, \bibinfo {author} {\bibfnamefont {G.}~\bibnamefont {Lani}},
  \bibinfo {author} {\bibfnamefont {F.}~\bibnamefont {Sottile}}, \bibinfo
  {author} {\bibfnamefont {P.}~\bibnamefont {Romaniello}}, \bibinfo {author}
  {\bibfnamefont {M.}~\bibnamefont {Gatti}}, \bibinfo {author} {\bibfnamefont
  {J.~J.}\ \bibnamefont {Kas}}, \bibinfo {author} {\bibfnamefont {J.~J.}\
  \bibnamefont {Rehr}}, \bibinfo {author} {\bibfnamefont {M.~G.}\ \bibnamefont
  {Silly}}, \bibinfo {author} {\bibfnamefont {F.}~\bibnamefont {Sirotti}}, \
  and\ \bibinfo {author} {\bibfnamefont {L.}~\bibnamefont {Reining}},\ }\href
  {\doibase 10.1103/PhysRevLett.107.166401} {\bibfield  {journal} {\bibinfo
  {journal} {Phys. Rev. Lett.}\ }\textbf {\bibinfo {volume} {107}},\ \bibinfo
  {pages} {166401} (\bibinfo {year} {2011})}\BibitemShut {NoStop}%
\bibitem [{\citenamefont {Kas}\ \emph {et~al.}(2014)\citenamefont {Kas},
  \citenamefont {Rehr},\ and\ \citenamefont {Reining}}]{kas:2014}%
  \BibitemOpen
  \bibfield  {author} {\bibinfo {author} {\bibfnamefont {J.~J.}\ \bibnamefont
  {Kas}}, \bibinfo {author} {\bibfnamefont {J.~J.}\ \bibnamefont {Rehr}}, \
  and\ \bibinfo {author} {\bibfnamefont {L.}~\bibnamefont {Reining}},\ }\href
  {\doibase 10.1103/PhysRevB.90.085112} {\bibfield  {journal} {\bibinfo
  {journal} {Phys. Rev. B}\ }\textbf {\bibinfo {volume} {90}},\ \bibinfo
  {pages} {085112} (\bibinfo {year} {2014})}\BibitemShut {NoStop}%
\bibitem [{\citenamefont {Caruso}\ \emph {et~al.}(2015)\citenamefont {Caruso},
  \citenamefont {Lambert},\ and\ \citenamefont {Giustino}}]{Caruso2015}%
  \BibitemOpen
  \bibfield  {author} {\bibinfo {author} {\bibfnamefont {F.}~\bibnamefont
  {Caruso}}, \bibinfo {author} {\bibfnamefont {H.}~\bibnamefont {Lambert}}, \
  and\ \bibinfo {author} {\bibfnamefont {F.}~\bibnamefont {Giustino}},\ }\href
  {\doibase 10.1103/PhysRevLett.114.146404} {\bibfield  {journal} {\bibinfo
  {journal} {Phys. Rev. Lett.}\ }\textbf {\bibinfo {volume} {114}},\ \bibinfo
  {pages} {146404} (\bibinfo {year} {2015})}\BibitemShut {NoStop}%
\bibitem [{\citenamefont {Gumhalter}\ \emph {et~al.}(2016)\citenamefont
  {Gumhalter}, \citenamefont {Kovac}, \citenamefont {Caruso}, \citenamefont
  {Lambert},\ and\ \citenamefont {Giustino}}]{gumhalter:2016}%
  \BibitemOpen
  \bibfield  {author} {\bibinfo {author} {\bibfnamefont {B.}~\bibnamefont
  {Gumhalter}}, \bibinfo {author} {\bibfnamefont {V.}~\bibnamefont {Kovac}},
  \bibinfo {author} {\bibfnamefont {F.}~\bibnamefont {Caruso}}, \bibinfo
  {author} {\bibfnamefont {H.}~\bibnamefont {Lambert}}, \ and\ \bibinfo
  {author} {\bibfnamefont {F.}~\bibnamefont {Giustino}},\ }\href {\doibase
  10.1103/PhysRevB.94.035103} {\bibfield  {journal} {\bibinfo  {journal} {Phys.
  Rev. B}\ }\textbf {\bibinfo {volume} {94}},\ \bibinfo {pages} {035103}
  (\bibinfo {year} {2016})}\BibitemShut {NoStop}%
\bibitem [{\citenamefont {Kutepov}(2016)}]{kutepov:2016}%
  \BibitemOpen
  \bibfield  {author} {\bibinfo {author} {\bibfnamefont {A.~L.}\ \bibnamefont
  {Kutepov}},\ }\href {\doibase 10.1103/PhysRevB.94.155101} {\bibfield
  {journal} {\bibinfo  {journal} {Phys. Rev. B}\ }\textbf {\bibinfo {volume}
  {94}},\ \bibinfo {pages} {155101} (\bibinfo {year} {2016})}\BibitemShut
  {NoStop}%
\bibitem [{\citenamefont {Kutepov}(2017)}]{kutepov:2017}%
  \BibitemOpen
  \bibfield  {author} {\bibinfo {author} {\bibfnamefont {A.~L.}\ \bibnamefont
  {Kutepov}},\ }\href {\doibase 10.1103/PhysRevB.95.195120} {\bibfield
  {journal} {\bibinfo  {journal} {Phys. Rev. B}\ }\textbf {\bibinfo {volume}
  {95}},\ \bibinfo {pages} {195120} (\bibinfo {year} {2017})}\BibitemShut
  {NoStop}%
\bibitem [{\citenamefont {Gr\"{u}neis}\ \emph {et~al.}(2014)\citenamefont
  {Gr\"{u}neis}, \citenamefont {Kresse}, \citenamefont {Hinuma},\ and\
  \citenamefont {Oba}}]{grueneis:2014}%
  \BibitemOpen
  \bibfield  {author} {\bibinfo {author} {\bibfnamefont {A.}~\bibnamefont
  {Gr\"{u}neis}}, \bibinfo {author} {\bibfnamefont {G.}~\bibnamefont {Kresse}},
  \bibinfo {author} {\bibfnamefont {Y.}~\bibnamefont {Hinuma}}, \ and\ \bibinfo
  {author} {\bibfnamefont {F.}~\bibnamefont {Oba}},\ }\href {\doibase
  10.1103/PhysRevLett.112.096401} {\bibfield  {journal} {\bibinfo  {journal}
  {Phys. Rev. Lett.}\ }\textbf {\bibinfo {volume} {112}},\ \bibinfo {pages}
  {096401} (\bibinfo {year} {2014})}\BibitemShut {NoStop}%
\bibitem [{\citenamefont {Maggio}\ and\ \citenamefont
  {Kresse}(2017)}]{maggio:2017a}%
  \BibitemOpen
  \bibfield  {author} {\bibinfo {author} {\bibfnamefont {E.}~\bibnamefont
  {Maggio}}\ and\ \bibinfo {author} {\bibfnamefont {G.}~\bibnamefont
  {Kresse}},\ }\href {\doibase 10.1021/acs.jctc.7b00586} {\bibfield  {journal}
  {\bibinfo  {journal} {Journal of Chemical Theory and Computation}\ }\textbf
  {\bibinfo {volume} {13}},\ \bibinfo {pages} {4765} (\bibinfo {year}
  {2017})}\BibitemShut {NoStop}%
\bibitem [{\citenamefont {McClain}\ \emph {et~al.}(2017)\citenamefont
  {McClain}, \citenamefont {Sun}, \citenamefont {Chan},\ and\ \citenamefont
  {Berkelbach}}]{mcclain:2017}%
  \BibitemOpen
  \bibfield  {author} {\bibinfo {author} {\bibfnamefont {J.}~\bibnamefont
  {McClain}}, \bibinfo {author} {\bibfnamefont {Q.}~\bibnamefont {Sun}},
  \bibinfo {author} {\bibfnamefont {G.~K.-L.}\ \bibnamefont {Chan}}, \ and\
  \bibinfo {author} {\bibfnamefont {T.~C.}\ \bibnamefont {Berkelbach}},\ }\href
  {\doibase 10.1021/acs.jctc.7b00049} {\bibfield  {journal} {\bibinfo
  {journal} {Journal of Chemical Theory and Computation}\ }\textbf {\bibinfo
  {volume} {13}},\ \bibinfo {pages} {1209} (\bibinfo {year}
  {2017})}\BibitemShut {NoStop}%
\bibitem [{\citenamefont {Baym}\ and\ \citenamefont
  {Kadanoff}(1961)}]{Baym1961}%
  \BibitemOpen
  \bibfield  {author} {\bibinfo {author} {\bibfnamefont {G.}~\bibnamefont
  {Baym}}\ and\ \bibinfo {author} {\bibfnamefont {L.~P.}\ \bibnamefont
  {Kadanoff}},\ }\href {\doibase 10.1103/PhysRev.124.287} {\bibfield  {journal}
  {\bibinfo  {journal} {Phys. Rev.}\ }\textbf {\bibinfo {volume} {124}},\
  \bibinfo {pages} {287} (\bibinfo {year} {1961})}\BibitemShut {NoStop}%
\bibitem [{\citenamefont {Baym}(1962)}]{Baym1962}%
  \BibitemOpen
  \bibfield  {author} {\bibinfo {author} {\bibfnamefont {G.}~\bibnamefont
  {Baym}},\ }\href {\doibase 10.1103/PhysRev.127.1391} {\bibfield  {journal}
  {\bibinfo  {journal} {Phys. Rev.}\ }\textbf {\bibinfo {volume} {127}},\
  \bibinfo {pages} {1391} (\bibinfo {year} {1962})}\BibitemShut {NoStop}%
\bibitem [{\citenamefont {Fuchs}\ \emph {et~al.}(2007)\citenamefont {Fuchs},
  \citenamefont {Furthm\"{u}ller}, \citenamefont {Bechstedt}, \citenamefont
  {Shishkin},\ and\ \citenamefont {Kresse}}]{fuchs:2007}%
  \BibitemOpen
  \bibfield  {author} {\bibinfo {author} {\bibfnamefont {F.}~\bibnamefont
  {Fuchs}}, \bibinfo {author} {\bibfnamefont {J.}~\bibnamefont
  {Furthm\"{u}ller}}, \bibinfo {author} {\bibfnamefont {F.}~\bibnamefont
  {Bechstedt}}, \bibinfo {author} {\bibfnamefont {M.}~\bibnamefont {Shishkin}},
  \ and\ \bibinfo {author} {\bibfnamefont {G.}~\bibnamefont {Kresse}},\ }\href
  {\doibase 10.1103/PhysRevB.76.115109} {\bibfield  {journal} {\bibinfo
  {journal} {Phys. Rev. B}\ }\textbf {\bibinfo {volume} {76}},\ \bibinfo
  {pages} {115109} (\bibinfo {year} {2007})}\BibitemShut {NoStop}%
\bibitem [{\citenamefont {Bruneval}\ and\ \citenamefont
  {Marques}(2013)}]{bruneval:2013}%
  \BibitemOpen
  \bibfield  {author} {\bibinfo {author} {\bibfnamefont {F.}~\bibnamefont
  {Bruneval}}\ and\ \bibinfo {author} {\bibfnamefont {M.~A.~L.}\ \bibnamefont
  {Marques}},\ }\href {\doibase 10.1021/ct300835h} {\bibfield  {journal}
  {\bibinfo  {journal} {Journal of Chemical Theory and Computation}\ }\textbf
  {\bibinfo {volume} {9}},\ \bibinfo {pages} {324} (\bibinfo {year}
  {2013})}\BibitemShut {NoStop}%
\bibitem [{\citenamefont {Zhu}\ and\ \citenamefont {Louie}(1991)}]{Zhu1991}%
  \BibitemOpen
  \bibfield  {author} {\bibinfo {author} {\bibfnamefont {X.}~\bibnamefont
  {Zhu}}\ and\ \bibinfo {author} {\bibfnamefont {S.~G.}\ \bibnamefont
  {Louie}},\ }\href {\doibase 10.1103/PhysRevB.43.14142} {\bibfield  {journal}
  {\bibinfo  {journal} {Phys. Rev. B}\ }\textbf {\bibinfo {volume} {43}},\
  \bibinfo {pages} {14142} (\bibinfo {year} {1991})}\BibitemShut {NoStop}%
\bibitem [{\citenamefont {Zakharov}\ \emph {et~al.}(1994)\citenamefont
  {Zakharov}, \citenamefont {Rubio}, \citenamefont {Blase}, \citenamefont
  {Cohen},\ and\ \citenamefont {Louie}}]{Zakharov1994}%
  \BibitemOpen
  \bibfield  {author} {\bibinfo {author} {\bibfnamefont {O.}~\bibnamefont
  {Zakharov}}, \bibinfo {author} {\bibfnamefont {A.}~\bibnamefont {Rubio}},
  \bibinfo {author} {\bibfnamefont {X.}~\bibnamefont {Blase}}, \bibinfo
  {author} {\bibfnamefont {M.~L.}\ \bibnamefont {Cohen}}, \ and\ \bibinfo
  {author} {\bibfnamefont {S.~G.}\ \bibnamefont {Louie}},\ }\href {\doibase
  10.1103/PhysRevB.50.10780} {\bibfield  {journal} {\bibinfo  {journal} {Phys.
  Rev. B}\ }\textbf {\bibinfo {volume} {50}},\ \bibinfo {pages} {10780}
  (\bibinfo {year} {1994})}\BibitemShut {NoStop}%
\bibitem [{\citenamefont {Shishkin}\ and\ \citenamefont
  {Kresse}(2007)}]{Shishkin_PRB2007}%
  \BibitemOpen
  \bibfield  {author} {\bibinfo {author} {\bibfnamefont {M.}~\bibnamefont
  {Shishkin}}\ and\ \bibinfo {author} {\bibfnamefont {G.}~\bibnamefont
  {Kresse}},\ }\href {\doibase 10.1103/PhysRevB.75.235102} {\bibfield
  {journal} {\bibinfo  {journal} {Phys. Rev. B}\ }\textbf {\bibinfo {volume}
  {75}},\ \bibinfo {pages} {235102} (\bibinfo {year} {2007})}\BibitemShut
  {NoStop}%
\bibitem [{\citenamefont {Faleev}\ \emph {et~al.}(2004)\citenamefont {Faleev},
  \citenamefont {van Schilfgaarde},\ and\ \citenamefont {Kotani}}]{Faleev2004}%
  \BibitemOpen
  \bibfield  {author} {\bibinfo {author} {\bibfnamefont {S.~V.}\ \bibnamefont
  {Faleev}}, \bibinfo {author} {\bibfnamefont {M.}~\bibnamefont {van
  Schilfgaarde}}, \ and\ \bibinfo {author} {\bibfnamefont {T.}~\bibnamefont
  {Kotani}},\ }\href {\doibase 10.1103/PhysRevLett.93.126406} {\bibfield
  {journal} {\bibinfo  {journal} {Phys. Rev. Lett.}\ }\textbf {\bibinfo
  {volume} {93}},\ \bibinfo {pages} {126406} (\bibinfo {year}
  {2004})}\BibitemShut {NoStop}%
\bibitem [{\citenamefont {van Schilfgaarde}\ \emph {et~al.}(2006)\citenamefont
  {van Schilfgaarde}, \citenamefont {Kotani},\ and\ \citenamefont
  {Faleev}}]{schilfgaarde:2006}%
  \BibitemOpen
  \bibfield  {author} {\bibinfo {author} {\bibfnamefont {M.}~\bibnamefont {van
  Schilfgaarde}}, \bibinfo {author} {\bibfnamefont {T.}~\bibnamefont {Kotani}},
  \ and\ \bibinfo {author} {\bibfnamefont {S.}~\bibnamefont {Faleev}},\ }\href
  {\doibase 10.1103/PhysRevLett.96.226402} {\bibfield  {journal} {\bibinfo
  {journal} {Phys. Rev. Lett.}\ }\textbf {\bibinfo {volume} {96}},\ \bibinfo
  {pages} {226402} (\bibinfo {year} {2006})}\BibitemShut {NoStop}%
\bibitem [{\citenamefont {Bruneval}\ \emph {et~al.}(2006)\citenamefont
  {Bruneval}, \citenamefont {Vast},\ and\ \citenamefont
  {Reining}}]{bruneval:2006}%
  \BibitemOpen
  \bibfield  {author} {\bibinfo {author} {\bibfnamefont {F.}~\bibnamefont
  {Bruneval}}, \bibinfo {author} {\bibfnamefont {N.}~\bibnamefont {Vast}}, \
  and\ \bibinfo {author} {\bibfnamefont {L.}~\bibnamefont {Reining}},\ }\href
  {\doibase 10.1103/PhysRevB.74.045102} {\bibfield  {journal} {\bibinfo
  {journal} {Phys. Rev. B}\ }\textbf {\bibinfo {volume} {74}},\ \bibinfo
  {pages} {045102} (\bibinfo {year} {2006})}\BibitemShut {NoStop}%
\bibitem [{\citenamefont {Shishkin}\ \emph {et~al.}(2007)\citenamefont
  {Shishkin}, \citenamefont {Marsman},\ and\ \citenamefont
  {Kresse}}]{shishkin:2007a}%
  \BibitemOpen
  \bibfield  {author} {\bibinfo {author} {\bibfnamefont {M.}~\bibnamefont
  {Shishkin}}, \bibinfo {author} {\bibfnamefont {M.}~\bibnamefont {Marsman}}, \
  and\ \bibinfo {author} {\bibfnamefont {G.}~\bibnamefont {Kresse}},\ }\href
  {\doibase 10.1103/PhysRevLett.99.246403} {\bibfield  {journal} {\bibinfo
  {journal} {Phys. Rev. Lett.}\ }\textbf {\bibinfo {volume} {99}},\ \bibinfo
  {pages} {246403} (\bibinfo {year} {2007})}\BibitemShut {NoStop}%
\bibitem [{\citenamefont {Bhandari}\ \emph {et~al.}(2018)\citenamefont
  {Bhandari}, \citenamefont {van Schilgaarde}, \citenamefont {Kotani},\ and\
  \citenamefont {Lambrecht}}]{Bhandari2018}%
  \BibitemOpen
  \bibfield  {author} {\bibinfo {author} {\bibfnamefont {C.}~\bibnamefont
  {Bhandari}}, \bibinfo {author} {\bibfnamefont {M.}~\bibnamefont {van
  Schilgaarde}}, \bibinfo {author} {\bibfnamefont {T.}~\bibnamefont {Kotani}},
  \ and\ \bibinfo {author} {\bibfnamefont {W.~R.~L.}\ \bibnamefont
  {Lambrecht}},\ }\href {\doibase 10.1103/PhysRevMaterials.2.013807} {\bibfield
   {journal} {\bibinfo  {journal} {Phys. Rev. Materials}\ }\textbf {\bibinfo
  {volume} {2}},\ \bibinfo {pages} {013807} (\bibinfo {year}
  {2018})}\BibitemShut {NoStop}%
\bibitem [{\citenamefont {Holm}\ and\ \citenamefont {von
  Barth}(1998)}]{Holm-scGW1998}%
  \BibitemOpen
  \bibfield  {author} {\bibinfo {author} {\bibfnamefont {B.}~\bibnamefont
  {Holm}}\ and\ \bibinfo {author} {\bibfnamefont {U.}~\bibnamefont {von
  Barth}},\ }\href {\doibase 10.1103/PhysRevB.57.2108} {\bibfield  {journal}
  {\bibinfo  {journal} {Phys. Rev. B}\ }\textbf {\bibinfo {volume} {57}},\
  \bibinfo {pages} {2108} (\bibinfo {year} {1998})}\BibitemShut {NoStop}%
\bibitem [{\citenamefont {Sch\"{o}ne}\ and\ \citenamefont
  {Eguiluz}(1998)}]{schoene:1998}%
  \BibitemOpen
  \bibfield  {author} {\bibinfo {author} {\bibfnamefont {W.-D.}\ \bibnamefont
  {Sch\"{o}ne}}\ and\ \bibinfo {author} {\bibfnamefont {A.~G.}\ \bibnamefont
  {Eguiluz}},\ }\href {\doibase 10.1103/PhysRevLett.81.1662} {\bibfield
  {journal} {\bibinfo  {journal} {Phys. Rev. Lett.}\ }\textbf {\bibinfo
  {volume} {81}},\ \bibinfo {pages} {1662} (\bibinfo {year}
  {1998})}\BibitemShut {NoStop}%
\bibitem [{\citenamefont {Stan}\ \emph {et~al.}(2009)\citenamefont {Stan},
  \citenamefont {Dahlen},\ and\ \citenamefont {van Leeuwen}}]{Adrian2009}%
  \BibitemOpen
  \bibfield  {author} {\bibinfo {author} {\bibfnamefont {A.}~\bibnamefont
  {Stan}}, \bibinfo {author} {\bibfnamefont {N.~E.}\ \bibnamefont {Dahlen}}, \
  and\ \bibinfo {author} {\bibfnamefont {R.}~\bibnamefont {van Leeuwen}},\
  }\href {\doibase 10.1063/1.3089567} {\bibfield  {journal} {\bibinfo
  {journal} {The Journal of Chemical Physics}\ }\textbf {\bibinfo {volume}
  {130}},\ \bibinfo {pages} {114105} (\bibinfo {year} {2009})}\BibitemShut
  {NoStop}%
\bibitem [{\citenamefont {Caruso}\ \emph {et~al.}(2012)\citenamefont {Caruso},
  \citenamefont {Rinke}, \citenamefont {Ren}, \citenamefont {Scheffler},\ and\
  \citenamefont {Rubio}}]{caruso:2012}%
  \BibitemOpen
  \bibfield  {author} {\bibinfo {author} {\bibfnamefont {F.}~\bibnamefont
  {Caruso}}, \bibinfo {author} {\bibfnamefont {P.}~\bibnamefont {Rinke}},
  \bibinfo {author} {\bibfnamefont {X.}~\bibnamefont {Ren}}, \bibinfo {author}
  {\bibfnamefont {M.}~\bibnamefont {Scheffler}}, \ and\ \bibinfo {author}
  {\bibfnamefont {A.}~\bibnamefont {Rubio}},\ }\href {\doibase
  10.1103/PhysRevB.86.081102} {\bibfield  {journal} {\bibinfo  {journal} {Phys.
  Rev. B}\ }\textbf {\bibinfo {volume} {86}},\ \bibinfo {pages} {081102}
  (\bibinfo {year} {2012})}\BibitemShut {NoStop}%
\bibitem [{\citenamefont {Kutepov}\ \emph {et~al.}(2012)\citenamefont
  {Kutepov}, \citenamefont {Haule}, \citenamefont {Savrasov},\ and\
  \citenamefont {Kotliar}}]{Kutepov2012}%
  \BibitemOpen
  \bibfield  {author} {\bibinfo {author} {\bibfnamefont {A.}~\bibnamefont
  {Kutepov}}, \bibinfo {author} {\bibfnamefont {K.}~\bibnamefont {Haule}},
  \bibinfo {author} {\bibfnamefont {S.~Y.}\ \bibnamefont {Savrasov}}, \ and\
  \bibinfo {author} {\bibfnamefont {G.}~\bibnamefont {Kotliar}},\ }\href
  {\doibase 10.1103/PhysRevB.85.155129} {\bibfield  {journal} {\bibinfo
  {journal} {Phys. Rev. B}\ }\textbf {\bibinfo {volume} {85}},\ \bibinfo
  {pages} {155129} (\bibinfo {year} {2012})}\BibitemShut {NoStop}%
\bibitem [{\citenamefont {Koval}\ \emph {et~al.}(2014)\citenamefont {Koval},
  \citenamefont {Foerster},\ and\ \citenamefont
  {S\'anchez-Portal}}]{Koval2014}%
  \BibitemOpen
  \bibfield  {author} {\bibinfo {author} {\bibfnamefont {P.}~\bibnamefont
  {Koval}}, \bibinfo {author} {\bibfnamefont {D.}~\bibnamefont {Foerster}}, \
  and\ \bibinfo {author} {\bibfnamefont {D.}~\bibnamefont {S\'anchez-Portal}},\
  }\href {\doibase 10.1103/PhysRevB.89.155417} {\bibfield  {journal} {\bibinfo
  {journal} {Phys. Rev. B}\ }\textbf {\bibinfo {volume} {89}},\ \bibinfo
  {pages} {155417} (\bibinfo {year} {2014})}\BibitemShut {NoStop}%
\bibitem [{\citenamefont {Gulans}(2014)}]{Gulans2014}%
  \BibitemOpen
  \bibfield  {author} {\bibinfo {author} {\bibfnamefont {A.}~\bibnamefont
  {Gulans}},\ }\href {\doibase 10.1063/1.4900447} {\bibfield  {journal}
  {\bibinfo  {journal} {The Journal of Chemical Physics}\ }\textbf {\bibinfo
  {volume} {141}},\ \bibinfo {pages} {164127} (\bibinfo {year}
  {2014})}\BibitemShut {NoStop}%
\bibitem [{\citenamefont {Klime\v{s}}\ \emph {et~al.}(2014)\citenamefont
  {Klime\v{s}}, \citenamefont {Kaltak},\ and\ \citenamefont
  {Kresse}}]{klimes:2014}%
  \BibitemOpen
  \bibfield  {author} {\bibinfo {author} {\bibfnamefont {J.}~\bibnamefont
  {Klime\v{s}}}, \bibinfo {author} {\bibfnamefont {M.}~\bibnamefont {Kaltak}},
  \ and\ \bibinfo {author} {\bibfnamefont {G.}~\bibnamefont {Kresse}},\ }\href
  {\doibase 10.1103/PhysRevB.90.075125} {\bibfield  {journal} {\bibinfo
  {journal} {Phys. Rev. B}\ }\textbf {\bibinfo {volume} {90}},\ \bibinfo
  {pages} {075125} (\bibinfo {year} {2014})}\BibitemShut {NoStop}%
\bibitem [{\citenamefont {Erg\"onenc}\ \emph {et~al.}(2018)\citenamefont
  {Erg\"onenc}, \citenamefont {Kim}, \citenamefont {Liu}, \citenamefont
  {Kresse},\ and\ \citenamefont {Franchini}}]{ergoenenc:2018}%
  \BibitemOpen
  \bibfield  {author} {\bibinfo {author} {\bibfnamefont {Z.}~\bibnamefont
  {Erg\"onenc}}, \bibinfo {author} {\bibfnamefont {B.}~\bibnamefont {Kim}},
  \bibinfo {author} {\bibfnamefont {P.}~\bibnamefont {Liu}}, \bibinfo {author}
  {\bibfnamefont {G.}~\bibnamefont {Kresse}}, \ and\ \bibinfo {author}
  {\bibfnamefont {C.}~\bibnamefont {Franchini}},\ }\href {\doibase
  10.1103/PhysRevMaterials.2.024601} {\bibfield  {journal} {\bibinfo  {journal}
  {Phys. Rev. Materials}\ }\textbf {\bibinfo {volume} {2}},\ \bibinfo {pages}
  {024601} (\bibinfo {year} {2018})}\BibitemShut {NoStop}%
\bibitem [{\citenamefont {Baroni}\ and\ \citenamefont
  {Resta}(1986)}]{Baroni1986}%
  \BibitemOpen
  \bibfield  {author} {\bibinfo {author} {\bibfnamefont {S.}~\bibnamefont
  {Baroni}}\ and\ \bibinfo {author} {\bibfnamefont {R.}~\bibnamefont {Resta}},\
  }\href {\doibase 10.1103/PhysRevB.33.7017} {\bibfield  {journal} {\bibinfo
  {journal} {Phys. Rev. B}\ }\textbf {\bibinfo {volume} {33}},\ \bibinfo
  {pages} {7017} (\bibinfo {year} {1986})}\BibitemShut {NoStop}%
\bibitem [{\citenamefont {Gajdo\v{s}}\ \emph {et~al.}(2006)\citenamefont
  {Gajdo\v{s}}, \citenamefont {Hummer}, \citenamefont {Kresse}, \citenamefont
  {Furthm\"{u}ller},\ and\ \citenamefont {Bechstedt}}]{gajdos:2006}%
  \BibitemOpen
  \bibfield  {author} {\bibinfo {author} {\bibfnamefont {M.}~\bibnamefont
  {Gajdo\v{s}}}, \bibinfo {author} {\bibfnamefont {K.}~\bibnamefont {Hummer}},
  \bibinfo {author} {\bibfnamefont {G.}~\bibnamefont {Kresse}}, \bibinfo
  {author} {\bibfnamefont {J.}~\bibnamefont {Furthm\"{u}ller}}, \ and\ \bibinfo
  {author} {\bibfnamefont {F.}~\bibnamefont {Bechstedt}},\ }\href {\doibase
  10.1103/PhysRevB.73.045112} {\bibfield  {journal} {\bibinfo  {journal} {Phys.
  Rev. B}\ }\textbf {\bibinfo {volume} {73}},\ \bibinfo {pages} {045112}
  (\bibinfo {year} {2006})}\BibitemShut {NoStop}%
\bibitem [{\citenamefont {Harl}\ \emph {et~al.}(2010)\citenamefont {Harl},
  \citenamefont {Schimka},\ and\ \citenamefont {Kresse}}]{harl:2010}%
  \BibitemOpen
  \bibfield  {author} {\bibinfo {author} {\bibfnamefont {J.}~\bibnamefont
  {Harl}}, \bibinfo {author} {\bibfnamefont {L.}~\bibnamefont {Schimka}}, \
  and\ \bibinfo {author} {\bibfnamefont {G.}~\bibnamefont {Kresse}},\ }\href
  {\doibase 10.1103/PhysRevB.81.115126} {\bibfield  {journal} {\bibinfo
  {journal} {Phys. Rev. B}\ }\textbf {\bibinfo {volume} {81}},\ \bibinfo
  {pages} {115126} (\bibinfo {year} {2010})}\BibitemShut {NoStop}%
\bibitem [{\citenamefont {Liu}\ \emph {et~al.}(2016)\citenamefont {Liu},
  \citenamefont {Kaltak}, \citenamefont {Klime\v{s}},\ and\ \citenamefont
  {Kresse}}]{liu:2016}%
  \BibitemOpen
  \bibfield  {author} {\bibinfo {author} {\bibfnamefont {P.}~\bibnamefont
  {Liu}}, \bibinfo {author} {\bibfnamefont {M.}~\bibnamefont {Kaltak}},
  \bibinfo {author} {\bibfnamefont {J.}~\bibnamefont {Klime\v{s}}}, \ and\
  \bibinfo {author} {\bibfnamefont {G.}~\bibnamefont {Kresse}},\ }\href
  {\doibase 10.1103/PhysRevB.94.165109} {\bibfield  {journal} {\bibinfo
  {journal} {Phys. Rev. B}\ }\textbf {\bibinfo {volume} {94}},\ \bibinfo
  {pages} {165109} (\bibinfo {year} {2016})}\BibitemShut {NoStop}%
\bibitem [{\citenamefont {Rojas}\ \emph {et~al.}(1995)\citenamefont {Rojas},
  \citenamefont {Godby},\ and\ \citenamefont {Needs}}]{Rojas1995}%
  \BibitemOpen
  \bibfield  {author} {\bibinfo {author} {\bibfnamefont {H.~N.}\ \bibnamefont
  {Rojas}}, \bibinfo {author} {\bibfnamefont {R.~W.}\ \bibnamefont {Godby}}, \
  and\ \bibinfo {author} {\bibfnamefont {R.~J.}\ \bibnamefont {Needs}},\ }\href
  {\doibase 10.1103/PhysRevLett.74.1827} {\bibfield  {journal} {\bibinfo
  {journal} {Phys. Rev. Lett.}\ }\textbf {\bibinfo {volume} {74}},\ \bibinfo
  {pages} {1827} (\bibinfo {year} {1995})}\BibitemShut {NoStop}%
\bibitem [{\citenamefont {Rieger}\ \emph {et~al.}(1999)\citenamefont {Rieger},
  \citenamefont {Steinbeck}, \citenamefont {White}, \citenamefont {Rojas},\
  and\ \citenamefont {Godby}}]{rieger:1999}%
  \BibitemOpen
  \bibfield  {author} {\bibinfo {author} {\bibfnamefont {M.~M.}\ \bibnamefont
  {Rieger}}, \bibinfo {author} {\bibfnamefont {L.}~\bibnamefont {Steinbeck}},
  \bibinfo {author} {\bibfnamefont {I.}~\bibnamefont {White}}, \bibinfo
  {author} {\bibfnamefont {H.}~\bibnamefont {Rojas}}, \ and\ \bibinfo {author}
  {\bibfnamefont {R.}~\bibnamefont {Godby}},\ }\href {\doibase
  https://doi.org/10.1016/S0010-4655(98)00174-X} {\bibfield  {journal}
  {\bibinfo  {journal} {Computer Physics Communications}\ }\textbf {\bibinfo
  {volume} {117}},\ \bibinfo {pages} {211} (\bibinfo {year}
  {1999})}\BibitemShut {NoStop}%
\bibitem [{\citenamefont {Kaltak}\ \emph
  {et~al.}(2014{\natexlab{a}})\citenamefont {Kaltak}, \citenamefont
  {Klime\v{s}},\ and\ \citenamefont {Kresse}}]{KaltakJCTC2014}%
  \BibitemOpen
  \bibfield  {author} {\bibinfo {author} {\bibfnamefont {M.}~\bibnamefont
  {Kaltak}}, \bibinfo {author} {\bibfnamefont {J.}~\bibnamefont {Klime\v{s}}},
  \ and\ \bibinfo {author} {\bibfnamefont {G.}~\bibnamefont {Kresse}},\ }\href
  {\doibase 10.1021/ct5001268} {\bibfield  {journal} {\bibinfo  {journal}
  {Journal of Chemical Theory and Computation}\ }\textbf {\bibinfo {volume}
  {10}},\ \bibinfo {pages} {2498} (\bibinfo {year}
  {2014}{\natexlab{a}})}\BibitemShut {NoStop}%
\bibitem [{\citenamefont {Kaltak}\ \emph
  {et~al.}(2014{\natexlab{b}})\citenamefont {Kaltak}, \citenamefont
  {Klime\v{s}},\ and\ \citenamefont {Kresse}}]{Kaltak2014}%
  \BibitemOpen
  \bibfield  {author} {\bibinfo {author} {\bibfnamefont {M.}~\bibnamefont
  {Kaltak}}, \bibinfo {author} {\bibfnamefont {J.}~\bibnamefont {Klime\v{s}}},
  \ and\ \bibinfo {author} {\bibfnamefont {G.}~\bibnamefont {Kresse}},\ }\href
  {\doibase 10.1103/PhysRevB.90.054115} {\bibfield  {journal} {\bibinfo
  {journal} {Phys. Rev. B}\ }\textbf {\bibinfo {volume} {90}},\ \bibinfo
  {pages} {054115} (\bibinfo {year} {2014}{\natexlab{b}})}\BibitemShut
  {NoStop}%
\bibitem [{\citenamefont {Maggio}\ \emph {et~al.}(2017)\citenamefont {Maggio},
  \citenamefont {Liu}, \citenamefont {van Setten},\ and\ \citenamefont
  {Kresse}}]{MaggioGW1002017}%
  \BibitemOpen
  \bibfield  {author} {\bibinfo {author} {\bibfnamefont {E.}~\bibnamefont
  {Maggio}}, \bibinfo {author} {\bibfnamefont {P.}~\bibnamefont {Liu}},
  \bibinfo {author} {\bibfnamefont {M.~J.}\ \bibnamefont {van Setten}}, \ and\
  \bibinfo {author} {\bibfnamefont {G.}~\bibnamefont {Kresse}},\ }\href
  {\doibase 10.1021/acs.jctc.6b01150} {\bibfield  {journal} {\bibinfo
  {journal} {Journal of Chemical Theory and Computation}\ }\textbf {\bibinfo
  {volume} {13}},\ \bibinfo {pages} {635} (\bibinfo {year} {2017})}\BibitemShut
  {NoStop}%
\bibitem [{\citenamefont {Tomczak}\ \emph {et~al.}(2017)\citenamefont
  {Tomczak}, \citenamefont {Liu}, \citenamefont {Toschi}, \citenamefont
  {Kresse},\ and\ \citenamefont {Held}}]{Tomczak2017}%
  \BibitemOpen
  \bibfield  {author} {\bibinfo {author} {\bibfnamefont {J.~M.}\ \bibnamefont
  {Tomczak}}, \bibinfo {author} {\bibfnamefont {P.}~\bibnamefont {Liu}},
  \bibinfo {author} {\bibfnamefont {A.}~\bibnamefont {Toschi}}, \bibinfo
  {author} {\bibfnamefont {G.}~\bibnamefont {Kresse}}, \ and\ \bibinfo {author}
  {\bibfnamefont {K.}~\bibnamefont {Held}},\ }\href {\doibase
  10.1140/epjst/e2017-70053-1} {\bibfield  {journal} {\bibinfo  {journal} {The
  European Physical Journal Special Topics}\ }\textbf {\bibinfo {volume}
  {226}},\ \bibinfo {pages} {2565} (\bibinfo {year} {2017})}\BibitemShut
  {NoStop}%
\bibitem [{\citenamefont {Wang}\ \emph {et~al.}(2011)\citenamefont {Wang},
  \citenamefont {Dang},\ and\ \citenamefont {Millis}}]{wang:2011}%
  \BibitemOpen
  \bibfield  {author} {\bibinfo {author} {\bibfnamefont {X.}~\bibnamefont
  {Wang}}, \bibinfo {author} {\bibfnamefont {H.~T.}\ \bibnamefont {Dang}}, \
  and\ \bibinfo {author} {\bibfnamefont {A.~J.}\ \bibnamefont {Millis}},\
  }\href {\doibase 10.1103/PhysRevB.84.073104} {\bibfield  {journal} {\bibinfo
  {journal} {Phys. Rev. B}\ }\textbf {\bibinfo {volume} {84}},\ \bibinfo
  {pages} {073104} (\bibinfo {year} {2011})}\BibitemShut {NoStop}%
\bibitem [{\citenamefont {Bl\"ochl}(1994)}]{BlochlPAW1994}%
  \BibitemOpen
  \bibfield  {author} {\bibinfo {author} {\bibfnamefont {P.~E.}\ \bibnamefont
  {Bl\"ochl}},\ }\href {\doibase 10.1103/PhysRevB.50.17953} {\bibfield
  {journal} {\bibinfo  {journal} {Phys. Rev. B}\ }\textbf {\bibinfo {volume}
  {50}},\ \bibinfo {pages} {17953} (\bibinfo {year} {1994})}\BibitemShut
  {NoStop}%
\bibitem [{\citenamefont {Kresse}\ and\ \citenamefont
  {Joubert}(1999)}]{KressePAW1999}%
  \BibitemOpen
  \bibfield  {author} {\bibinfo {author} {\bibfnamefont {G.}~\bibnamefont
  {Kresse}}\ and\ \bibinfo {author} {\bibfnamefont {D.}~\bibnamefont
  {Joubert}},\ }\href {\doibase 10.1103/PhysRevB.59.1758} {\bibfield  {journal}
  {\bibinfo  {journal} {Phys. Rev. B}\ }\textbf {\bibinfo {volume} {59}},\
  \bibinfo {pages} {1758} (\bibinfo {year} {1999})}\BibitemShut {NoStop}%
\bibitem [{\citenamefont {Baker}(1975)}]{pade1975}%
  \BibitemOpen
  \bibfield  {author} {\bibinfo {author} {\bibfnamefont {G.~A.~J.}\
  \bibnamefont {Baker}},\ }\href@noop {} {\emph {\bibinfo {title} {Essentials
  of Pad\'{e} Approximants}}}\ (\bibinfo  {publisher} {Academic Press, New
  York},\ \bibinfo {year} {1975})\ Chap.~\bibinfo {chapter} {18}\BibitemShut
  {NoStop}%
\bibitem [{not()}]{note}%
  \BibitemOpen
  \href@noop {} {\emph {\bibinfo {title} {\rm Note that for metals, $\chi^{\rm
  KS}_{0,0}(\qq,\ii\omega) = a_0+a \qq^2 + b \qq^4 + \mathcal{O}(\qq^6)$ with
  $a_0 \propto \beta$, where $\beta$ is the inverse of the
  temperature}}}\BibitemShut {NoStop}%
\bibitem [{\citenamefont {Kresse}\ and\ \citenamefont
  {Hafner}(1993)}]{KresseVASP1993}%
  \BibitemOpen
  \bibfield  {author} {\bibinfo {author} {\bibfnamefont {G.}~\bibnamefont
  {Kresse}}\ and\ \bibinfo {author} {\bibfnamefont {J.}~\bibnamefont
  {Hafner}},\ }\href {\doibase 10.1103/PhysRevB.47.558} {\bibfield  {journal}
  {\bibinfo  {journal} {Phys. Rev. B}\ }\textbf {\bibinfo {volume} {47}},\
  \bibinfo {pages} {558} (\bibinfo {year} {1993})}\BibitemShut {NoStop}%
\bibitem [{\citenamefont {Kresse}\ and\ \citenamefont
  {Furthm\"uller}(1996)}]{KresseVASP1996}%
  \BibitemOpen
  \bibfield  {author} {\bibinfo {author} {\bibfnamefont {G.}~\bibnamefont
  {Kresse}}\ and\ \bibinfo {author} {\bibfnamefont {J.}~\bibnamefont
  {Furthm\"uller}},\ }\href {\doibase 10.1103/PhysRevB.54.11169} {\bibfield
  {journal} {\bibinfo  {journal} {Phys. Rev. B}\ }\textbf {\bibinfo {volume}
  {54}},\ \bibinfo {pages} {11169} (\bibinfo {year} {1996})}\BibitemShut
  {NoStop}%
\bibitem [{\citenamefont {Perdew}\ \emph {et~al.}(1996)\citenamefont {Perdew},
  \citenamefont {Burke},\ and\ \citenamefont {Ernzerhof}}]{PBE1996}%
  \BibitemOpen
  \bibfield  {author} {\bibinfo {author} {\bibfnamefont {J.~P.}\ \bibnamefont
  {Perdew}}, \bibinfo {author} {\bibfnamefont {K.}~\bibnamefont {Burke}}, \
  and\ \bibinfo {author} {\bibfnamefont {M.}~\bibnamefont {Ernzerhof}},\ }\href
  {\doibase 10.1103/PhysRevLett.77.3865} {\bibfield  {journal} {\bibinfo
  {journal} {Phys. Rev. Lett.}\ }\textbf {\bibinfo {volume} {77}},\ \bibinfo
  {pages} {3865} (\bibinfo {year} {1996})}\BibitemShut {NoStop}%
\bibitem [{\citenamefont {Perdew}\ \emph {et~al.}(1997)\citenamefont {Perdew},
  \citenamefont {Burke},\ and\ \citenamefont {Ernzerhof}}]{Perdew1997}%
  \BibitemOpen
  \bibfield  {author} {\bibinfo {author} {\bibfnamefont {J.~P.}\ \bibnamefont
  {Perdew}}, \bibinfo {author} {\bibfnamefont {K.}~\bibnamefont {Burke}}, \
  and\ \bibinfo {author} {\bibfnamefont {M.}~\bibnamefont {Ernzerhof}},\ }\href
  {\doibase 10.1103/PhysRevLett.78.1396} {\bibfield  {journal} {\bibinfo
  {journal} {Phys. Rev. Lett.}\ }\textbf {\bibinfo {volume} {78}},\ \bibinfo
  {pages} {1396} (\bibinfo {year} {1997})}\BibitemShut {NoStop}%
\bibitem [{\citenamefont {Heyd}\ \emph {et~al.}(2003)\citenamefont {Heyd},
  \citenamefont {Scuseria},\ and\ \citenamefont {Ernzerhof}}]{hse:2003}%
  \BibitemOpen
  \bibfield  {author} {\bibinfo {author} {\bibfnamefont {J.}~\bibnamefont
  {Heyd}}, \bibinfo {author} {\bibfnamefont {G.~E.}\ \bibnamefont {Scuseria}},
  \ and\ \bibinfo {author} {\bibfnamefont {M.}~\bibnamefont {Ernzerhof}},\
  }\href {\doibase 10.1063/1.1564060} {\bibfield  {journal} {\bibinfo
  {journal} {The Journal of Chemical Physics}\ }\textbf {\bibinfo {volume}
  {118}},\ \bibinfo {pages} {8207} (\bibinfo {year} {2003})}\BibitemShut
  {NoStop}%
\bibitem [{\citenamefont {Paier}\ \emph {et~al.}(2005)\citenamefont {Paier},
  \citenamefont {Hirschl}, \citenamefont {Marsman},\ and\ \citenamefont
  {Kresse}}]{PaierHSE2005}%
  \BibitemOpen
  \bibfield  {author} {\bibinfo {author} {\bibfnamefont {J.}~\bibnamefont
  {Paier}}, \bibinfo {author} {\bibfnamefont {R.}~\bibnamefont {Hirschl}},
  \bibinfo {author} {\bibfnamefont {M.}~\bibnamefont {Marsman}}, \ and\
  \bibinfo {author} {\bibfnamefont {G.}~\bibnamefont {Kresse}},\ }\href
  {\doibase 10.1063/1.1926272} {\bibfield  {journal} {\bibinfo  {journal} {The
  Journal of Chemical Physics}\ }\textbf {\bibinfo {volume} {122}},\ \bibinfo
  {pages} {234102} (\bibinfo {year} {2005})}\BibitemShut {NoStop}%
\bibitem [{\citenamefont {Shih}\ \emph {et~al.}(2010)\citenamefont {Shih},
  \citenamefont {Xue}, \citenamefont {Zhang}, \citenamefont {Cohen},\ and\
  \citenamefont {Louie}}]{PhysRevLett.105.146401}%
  \BibitemOpen
  \bibfield  {author} {\bibinfo {author} {\bibfnamefont {B.-C.}\ \bibnamefont
  {Shih}}, \bibinfo {author} {\bibfnamefont {Y.}~\bibnamefont {Xue}}, \bibinfo
  {author} {\bibfnamefont {P.}~\bibnamefont {Zhang}}, \bibinfo {author}
  {\bibfnamefont {M.~L.}\ \bibnamefont {Cohen}}, \ and\ \bibinfo {author}
  {\bibfnamefont {S.~G.}\ \bibnamefont {Louie}},\ }\href {\doibase
  10.1103/PhysRevLett.105.146401} {\bibfield  {journal} {\bibinfo  {journal}
  {Phys. Rev. Lett.}\ }\textbf {\bibinfo {volume} {105}},\ \bibinfo {pages}
  {146401} (\bibinfo {year} {2010})}\BibitemShut {NoStop}%
\bibitem [{\citenamefont {Friedrich}\ \emph {et~al.}(2011)\citenamefont
  {Friedrich}, \citenamefont {M\"uller},\ and\ \citenamefont
  {Bl\"ugel}}]{PhysRevB.83.081101}%
  \BibitemOpen
  \bibfield  {author} {\bibinfo {author} {\bibfnamefont {C.}~\bibnamefont
  {Friedrich}}, \bibinfo {author} {\bibfnamefont {M.~C.}\ \bibnamefont
  {M\"uller}}, \ and\ \bibinfo {author} {\bibfnamefont {S.}~\bibnamefont
  {Bl\"ugel}},\ }\href {\doibase 10.1103/PhysRevB.83.081101} {\bibfield
  {journal} {\bibinfo  {journal} {Phys. Rev. B}\ }\textbf {\bibinfo {volume}
  {83}},\ \bibinfo {pages} {081101} (\bibinfo {year} {2011})}\BibitemShut
  {NoStop}%
\bibitem [{\citenamefont {Nunes}\ and\ \citenamefont
  {Gonze}(2001)}]{nunes:2001}%
  \BibitemOpen
  \bibfield  {author} {\bibinfo {author} {\bibfnamefont {R.~W.}\ \bibnamefont
  {Nunes}}\ and\ \bibinfo {author} {\bibfnamefont {X.}~\bibnamefont {Gonze}},\
  }\href {\doibase 10.1103/PhysRevB.63.155107} {\bibfield  {journal} {\bibinfo
  {journal} {Phys. Rev. B}\ }\textbf {\bibinfo {volume} {63}},\ \bibinfo
  {pages} {155107} (\bibinfo {year} {2001})}\BibitemShut {NoStop}%
\end{thebibliography}%

\end{document}